\documentclass[twocolumn]{aastex62}

\received{July 20, 2018}
\revised{February 6, 2019}
\submitjournal{ApJ}

\shorttitle{Optical counterparts of ULXs in NGC 4490/4485}
\shortauthors{Avdan et al.}

\begin{document}

\title{OPTICAL COUNTERPARTS OF ULXs AND THEIR HOST ENVIRONMENTS IN NGC 4490/4485}

\correspondingauthor{Senay Avdan}
\email{kayaci.s@gmail.com}

\author{Senay Avdan}
\affil{Space Sciences and Solar Energy Research and Application Center (UZAYMER), University of Cukurova, Adana, Turkey}

\author{Aysun Akyuz}
\affil{Department of Physics, University of Cukurova, Adana, Turkey}
\affil{Space Sciences and Solar Energy Research and Application Center (UZAYMER), University of Cukurova, Adana, Turkey}

\author{Alexander Vinokurov}
\affil{Special Astrophysical Observatory of the Russian AS, Nizhnij Arkhyz, Russia}

\author{Nazim Aksaker}
\affil{Adana Organized Industrial Zone Vocational School of Technical Sciences, Cukurova University, 01410 Adana, Turkey}
\affil{Space Sciences and Solar Energy Research and Application Center (UZAYMER), University of Cukurova, Adana, Turkey}

\author{Hasan Avdan}
\affil{Space Sciences and Solar Energy Research and Application Center (UZAYMER), University of Cukurova, Adana, Turkey}

\author{Sergei Fabrika}
\affil{Special Astrophysical Observatory of the Russian AS, Nizhnij Arkhyz, Russia}
\affil{Kazan Federal University, Kazan, Russia}

\author{Azamat Valeev}
\affil{Special Astrophysical Observatory of the Russian AS, Nizhnij Arkhyz, Russia}

\author{Inci Akkaya-Oralhan}
\affil{Department of Astronomy and Space Sciences, Erciyes University, 39039 Kayseri, Turkey}
\affil{Erciyes University, Astronomy and Space Science Observatory, 38039 Kayseri, Turkey}

\author{\c{S}\"{o}len Balman}
\affil{Currently, Self-Employed, Istanbul, 34381, Turkey}
\affil{Middle East Technical University, Dept. of Physics, Dumlupınar Bul., Univ. Mah. No.1, 06800, Ankara, Turkey}



\begin{abstract}

We report the identification of the possible optical counterparts of five out of seven Ultraluminous X-ray Sources (ULXs) in NGC 4490/4485 galaxy pair. Using archival Hubble Space Telescope ({\it HST}) imaging data, we identified a single optical candidate for two ULXs (X-4 and X-7) and multiple optical candidates for the other three ULXs (X-2, X-3 and X-6) within $\sim$ $0\farcs2$ error radius at the 90\% confidence level. Of the two remaining ULXs, X-1 has no {\it HST} imaging data and photometry could not be performed due to the position of X-5 in NGC4490. Absolute magnitudes ($M_{V}$) of the optical candidates lie between $-5.7$ and $-3.8$. Color-Magnitude Diagrams (CMDs) have been used to investigate the properties of counterparts and their environments. The locations of the counterparts of X-2, X-4, and X-6 suggest possible association with nearby group of stars while others have no association with a star cluster or group of stars. For comparison purposes, we analyzed previously unused three archival XMM-Newton observations. The long-term X-ray light curves of the sources (except transient X-7) show variability by a factor of three in a time scale more than a decade. The use of disk blackbody model for the mass of the compact objects indicates that these objects might have masses most likely in the range 10$-$15 $M_{\sun}$.
 
\end{abstract}



\section{Introduction} \label{sec:intro}

Ultraluminous X-ray sources (ULXs) are non-nuclear point-like sources in external galaxies with an X-ray luminosity exceeding the Eddington limit for a stellar-mass black hole ($L_{\mathrm{X}} \gtrsim 10^{39}$ erg s$^{-1}$). There are  few scenarios suggested to explain their high luminosities. According to two common ones, either accretors in ULXs are intermediate-mass black holes with standard accretion discs or stellar-mass black holes with super-critical discs. Recent studies promote stellar-mass black holes as accretors in ULXs (Liu et al. 2013; Motch et al. 2014; Fabrika et al. 2015). On the other hand, the recent discoveries of four ULXs that exhibited pulsed X-ray emission as expected from neutron stars, make their nature highly controversial (M82 X-2, Bachetti et al. 2014; NGC5907 ULX-1, Israel et al. 2017a; NGC 7793 P13, F\"{u}rst et al. 2016; Israel et al. 2017b and NGC 300 ULX1, Carpano et al. 2018). 

Multiwavelength observations of ULXs have played a key role to investigate their emission mechanisms and environments. Particularly, identification of the possible optical counterparts of ULXs and their broadband photometry allow us to constrain the mass and spectral type of the companion star.It is known that the best way to determine the mass of compact objects in ULXs is through dynamical mass measurements of the binary system. However, it is difficult to obtain the radial velocity curve and measure the mass function due to the fact that counterparts of ULXs are faint sources in the optical band ($m_{V}$ = 21-24) (Tao et al. 2011). Using the ground-based telescopes and {\it Hubble Space Telescope (HST)}, possible optical candidates of about 20 ULXs have been identified so far. Such studies may provide information about donor star (e.g. age, mass, and spectral type), host cluster and origin of optical emission (from accretion disc and/or donor star) (Kaaret et al. 2017).

Observations of the optical counterparts revealed that some of the ULXs are associated with young star clusters or star forming regions (Soria et al. 2005; Kaaret 2005; Abolmasov et al. 2007b; Grise et al. 2008, 2011, 2012; Avdan et al. 2016a,b). These findings have been supported by identifying point like counterparts of ULXs with blue colors which indicative of early type, OB stars (Liu et al. 2004; Grise et al. 2005; Soria et al. 2005; Roberts et al. 2008; Poutanen et al. 2013). These blue colors are consistent with emission from an irradiated accretion disk, however, it is noticed that the observed blue color may be partly due to contamination by the X-ray irradiation of the accretion disk and/or the companion star facing the X-ray source (Patruno \& Zampieri 2010; Grise et al. 2012; Jonker et al. 2012; Vinokurov et al. 2018). However, in only one case (P13 in NGC7793) photospheric absorption lines have been detected from the donor star on the blue part of spectrum (Motch et al. 2014). Furthermore, the association between young star clusters and ULXs implies that some of the donor stars can be red supergiants and these donors could be identified in the near-infrared band. The systematic search of counterparts in the near-infrared ({\it H} band) revealed that they are bright sources (Heida et al. 2014, 2016; Lopez et al. 2017). This could be important because the contribution of the accretion disk is lower in the near-infrared band than in the optical band. In addition, it is expected that the irradiation of donors of these binaries with large separations does not have a significant effect on observed emission (Copperwheat 2007; Heida et al. 2016).

Associations between ULXs and star clusters have also been studied in interacting galaxies. They are known to host a higher average number ($>5$) of ULXs. Therefore, they are good candidates to examine the population properties of ULXs. Poutanen et al. (2013) had extensively examined the significant associations between ULXs and stellar clusters in Antennae galaxies. Using {\it HST} and VLT data supplemented with theoretical stellar isochrones, they estimated the ages of these clusters as $<6$ Myr. It was discussed that these ULXs were probably ejected from the cluster in the evolutionary process; thus these sources might be high mass X-ray binaries instead of intermediate mass black holes. Another well-known interacting galaxy is NGC4490/NGC4485 at a distance of 7.8 Mpc (Tully 1988). NGC 4490 is a late-type spiral galaxy and NGC 4485 is an irregular galaxy. Their linear sizes are 15 kpc for NGC 4490 and 5.6 kpc for NGC 4485. Radio observations show that the star formation in NGC 4490 has been ongoing with constant rate of $\sim$4.7 $M_{\sun}$ yr$^{-1}$ (Clemens et al. 1999). Also NGC 4490 has a giant HI envelope that probably originated from the star formation (Clemens \& Alexander 2002). Our aim in this study is to identify the possible optical counterparts of ULXs in this galaxy pair and to investigate their associations with star groups or clusters. Previously, the three ULXs in this pair were detected by the {\it ROSAT} HRI observations (Roberts \& Warwick 2000). Later, using the {\it Chandra} ACIS-S observation, three more ULXs were identified by Roberts et al. (2002). The calculated unabsorbed luminosities of six ULXs (in the 0.5$-$8 keV band) fall into the range $(2.6-4.9)\times10^{39}$ erg s$^{-1}$. In addition, they noted that these ULXs appear to be spatially coincident with the star formation regions in the pair. Further, Fridriksson et al. (2008) searched long term variability of 38 X-ray sources in this galaxy pair using three {\it Chandra} observations. Eight of these sources were classified as ULXs in the luminosity range $(0.6-3)\times10^{39}$ erg s$^{-1}$. One of them is a transient ULX detected in a single observation (ID 4726). Gladstone \& Roberts (2009) investigated spectral and temporal features of seven ULXs (except ULX X-5, this source was ignored because of its low luminosity $L_{\mathrm{X}} \sim6$ $\times10^{38}$ erg s$^{-1}$ which was given in Table 5 of Fridriksson et al. 2008) using the same {\it Chandra} and {\it XMM-Newton} data sets. The $L_{\mathrm{X}}$ values of these sources are given in the range $(0.9-4)\times10^{39}$ erg s$^{-1}$ within the 0.5$-$8 keV energy band. The six of these seven sources (except the transient one which is X-7 in this paper) were classified as ULXs by Swartz et al. (2011). 

In the present work, the possible optical counterparts of these 7 ULXs in the galaxy pair NGC 4490/4485 have been searched extensively using archival {\it HST} and {\it XMM-Newton} data. The results from the analysis for the optically identified five ULXs (X-2, X-3, X-4, X-6 and X-7) are discussed. The optical spectra of environment of ULXs obtained by BTA 6-m telescope in SAO RAS are also examined. Three color Sloan Digital Sky Survey (SDSS) image of this galaxy pair with the approximate positions of seven ULXs (X1$-$7)\footnote{Sources are named from X-1 to X-6 in increasing R.A. order as given in Table 1 of Swartz et al. (2011) and X-7 is named from Gladstone \& Roberts (2009).} is given in Figure 1.

The paper is organized as follows: Data reduction and results are described in Section 2. The details and results of the optical analyses are given in subsections 2.1, 2.2 and 2.3. The X-ray observations are described in subsection 2.4. Discussion on the physical properties of the ULXs and conclusions are given in Section 3.

\section{Data Reduction and Results}

\subsection{{\it HST} Observations and Astrometry}

Identification of the optical candidates of the ULXs in the pair galaxies requires precise source positions. For this, an intercomparison of {\it Chandra}, {\it HST} and {\it SDSS} images were carried out to obtain improved astrometry. For the aimed astrometric correction, we have chosen deep {\it Chandra} Advanced CCD Imaging Spectrometer (ACIS) observation (ID4726) and two {\it HST} observations with Advanced Camera for Surveys (ACS, data set j9h807020) and Wide Field Camera 3 (WFC3, icdm42060). The counterpart of X-2 was located on ACS images while other counterparts were located on the WFC3 image. Logs of {\it HST} observations are given in Table 1.

Source detections in {\it Chandra} image were carried out using the {\scshape wavedetect} task in {\scshape ciao}. All ULXs are located on chip S3 of ACIS with a moderate offset to the optical axis from $1\farcm9$ to $2\farcm3$. Due to lack of the good X-ray/optical reference sources in the {\it HST} field, the offsets between {\it Chandra} and {\it HST} were corrected using the SDSS {\it r}-band image (Sloan Digital Sky Survey; Alam et al. 2015). Two relatively bright objects on the SDSS image were identified as the X-ray reference sources (given in Table 2). The most bright source s1 has 615 photons detected by {\it Chandra} with the statistical error radius of $0\farcs12$ at 90\% confidence level (offset to the optical axis is about $1\farcm8$). s2 has a large offset to the optical axis of $3\farcm3$, and for this source there only 62 photons detected by {\it Chandra}. Therefore, its position uncertainty at 90\% is about $0\farcs5$. Because of a big position error of s2 we derived offset between ACIS and SDSS images using only s1 (Table 2). Nevertheless we note that both reference sources give nearly the same offset values. 

Since the rotation of the X-ray image cannot be fixed with a single reference source, we used the estimation of rotation accuracy from Yang et al. (2011). From a few Chandra observations with multiple optically identified X-ray sources, the autors obtained a typical rotation about 2$\arcmin$ between Chandra and 2MASS. Using this value and distances between the ULXs and reference source s1 (about $200\arcsec$) the typical error caused by rotation is about $0\farcs1$. Combining in quadrature, these errors and uncertainties of ULX positions on ACIS image, also as uncertainties of s1 position both ACIS and SDSS images, we estimate that the corrected to SDSS ULX positions have uncertainties of about $0\farcs2$ at the 90\% confidence level (Table 2). On the other hand, using ten bright isolated reference stars on each {\it HST} images we derived shifts between SDSS and {\it HST} images with uncertainties better than $0\farcs04$ at 90\%. The corrected positions of the ULXs relative to {\it HST} are in Table 2. The main uncertainty is between {\it Chandra} and SDSS, all other errors are significantly smaller.

After astrometric correction, possible optical counterparts were identified for only 5 ULXs out of 7. While a single candidate were found for X-4, and X-7, more than one candidate were identified for X-3, X-2 and X-6 within their error circles. Photometry could not be performed for X-5 and X-1 because X-5 is located within the luminous part of the galaxy and X-1  has not been observed with {\it HST}.

The magnitudes of the counterpart candidates were calculated using the point-spread function (PSF) photometry. PSF was carried out with {\scshape dolphot} software (version 2.0, Dolphin 2000) for WFC3 and ACS data. Standard image reduction algorithms (bias and dark current subtraction, flat fielding) have been applied to FITS files were retrieved from {\it HST} data. The WFC3MASK, ACSMASK and SPLITGROUPS tasks were used to mask out all the bad pixels and split the multi-image FITS files into a single file per chip, respectively. Then the DOLPHOT task was used for source detection, photometry, and photometric conversion. The magnitudes were derived in the VEGA magnitude system used for WFC3 and ACS data. The magnitudes of the counterparts were corrected for the total extinction determined from ratios of Balmer lines of several nebulae around the candidates. Therefore, the spectral data obtained from 6-m BTA telescope were used to determine Balmer lines. 
 
\subsection{BTA Observations}

The ground-based optical spectral observations of the stellar association were made with the SCORPIO instrument at the 6-m BTA telescope of the SAO RAS (Afanasiev \& Moiseev 2005). The observations were carried out on 2011 January 4 (4800 s) for X-3 and X-6 and on 2016 March 13 (2400 s) for X-4. The VPHG550G grism (3500$-$7200 \AA) and 1$\arcsec$ slit were used for all observations where the seeing was about $\sim$(2-3) arcsec.

The standard data reduction steps were carried out with {\scshape iraf} software. Wavelength calibrations were done using Neon lamps. The spectral data of the standard stars taken the same night were used for flux calibration. The 2D spectra of the region were converted 1D spectra using {\scshape apall} task in {\scshape iraf}.

Strong emission lines such as $\mathrm{H}\beta$ ($\lambda$ 4861), [O {\scshape iii}] ($\lambda$ 4959,5007), $\mathrm{H}\alpha$ ($\lambda$ 6563) and [S {\scshape ii}] ($\lambda$ 6717+6731) were identified in whole spectral dataset. We estimated an average redshift value from these emission lines as z $\sim 0.0017$. This value is compatible with the redshift of the galaxy NGC 4490 ($z_{g}$ $\sim 0.0018$, Strauss et al. 1992).

Several nebulae were identified around the ULXs within the different slit positions having strong $\mathrm{H}\beta$, [O {\scshape iii}] and $\mathrm{H}\alpha$ lines in the spectra. The $\mathrm{H}\alpha$ and $\mathrm{H}\beta$ fluxes of the nebulae were measured to determine the observed Balmer decrement and corresponding reddening $E(\bv)$ value. To determine $E(\bv)$, the standard $(\mathrm{H}\alpha /\mathrm{H}\beta)_{\mathrm{int}} = 2.87$ value for star-forming galaxies was adopted that corresponds to electron density $n_{e} = 10^{2}$ cm$^{-3}$ and temperature $T= 10^{4}$ K for Case B of Osterbrock (1989) (Calzetti et al. 1994, 2001). Finally, these values were converted to $A_{V}$ using Cardelli law (Cardelli et al. 1989) in order to calculate for reddening correction. These $A_{V}$ values and the distances of ULXs to the nebulae are given in Table 3. In these calculations, the uncertainties in the distances are estimated in the range of $\sim 0\farcs5 - 0\farcs8$ (38 pc $/1\arcsec$).

\subsection{The Properties of Optical Counterparts}

In this section, the individual optical properties of the possible counterpart candidates are given in detail.

\textbf{X-2 :}
This source was identified in several datasets listed in Table 1. By carefully examining all the images, X-2 seems to have a faint structure compatible with an extended source within the $0\farcs15$ error radius (see Figure 2). At about $2\arcsec$ to the south$-$east of X-2, a small group of stars with an extent of $\sim$ 136 pc diameter is indicated by a circle on the F814W image in Figure 2. As a result of PSF photometry of X-2, three optical sources have been identified as counterpart candidates within the error circle. Previously the optical counterpart of X-2 was studied by Roberts et al. (2008) using only the year of 2005 {\it HST} data. They reported only one possible optical counterpart based on a $\sim$ $0\farcs6$ error on the astrometric precision of the X ray source position.

Dereddened magnitudes, color values of counterparts of X-2 obtained after photometric analysis are given in Table 4. To determine the ages and masses of counterpart candidates we built Color-Magnitude Diagrams (CMDs) using the PARSEC (PAdova and TRieste Stellar Evolution Code) isochrones (Bressan et al. 2012), on which extinction values ($A_{V}$ and $A_{I}$) from the calculated Balmer decrement (see Table 3) are overlaid using a distance modulus (DM) of 29.46 (for a distance of 7.8 Mpc, Tully 1988). These values are kept fixed to create the CMDs. The metallicity value for NGC 4490/4485 was taken as Z$=$0.005 (Esposito et al. 2013). One of the resultant CMDs, F814W versus F606W$-$814W, for possible counterparts of X-2 and nearby group of stars is given in Figure 3. The mean extinction along the direction of this source was calculated as $E(V-I)= 0.40$. The age and mass of candidate sources src1, src2 and src3 were obtained as $\sim$ 50, 80 and 90 Myr and $\sim$ 7, 6 and 5 $M_{\sun}$, respectively. Assuming the optical emission of donor star dominates, the spectral types of candidate counterparts were determined as B9$-$A2 supergiant for src1, G0 supergiant for src2 and A7 supergiant for src3 where intrinsic colors and the absolute magnitudes of Schmidt-Kaler table were used (Aller et al. 1982). The same assumption and procedures were followed to determine the spectral class of all counterpart candidates. In addition, only one of the CMDs created was shown due to the relatively large number of ULXs that have been examined.

The detected group of stars near X-2 does not indicate a homogeneous population. As seen in Figure 3, the CMD yielded diversity in the magnitude and age values of stars. However, the optical candidates of X-2 seem to be compatible with the oldest and faintest stars in this group. This may support the possibility that X-2 was ejected from the group.

\textbf{X-3 :}
After the astrometric correction, two faint candidates for the ULX X-3 were identified. These candidates are located near ($\sim 2 \arcsec$) the star cluster also catalogued as a super-star cluster (Figure 4) based on the initial mass function determination with the data obtained from the SDSS (Dowell et al. 2008). CMDs for these two sources and the nearby super-star cluster are also produced and a diagram F814W versus F555W$-$F814W is shown in Figure 5. Bright center of cluster may consist of multiple unresolved sources on all available images of this source. For this reason, a small number of cluster members could be selected by PSF photometry. The {\it HST} dereddened magnitudes, color, absolute magnitude of optical candidates are given in Table 5. Possible spectral type of optical candidates of X-3 seem to fall into the interval of F2$-$G8 supergiant for src1 and G0 supergiant for src2.

There is a cataloged super-star cluster $\sim 2\arcsec$ away from X-3. This would be a notable feature since the high stellar density within the superstar clusters are the most promising location for the formation of intermediate mass black holes via merging of massive stars (Kaaret et al. 2017). The core of the cluster is very bright therefore not many stars were identified using PSF photometry. The reddened magnitudes of the selected stars in the cluster are in the range of $m_{V} \sim 21-26$ but for the counterparts of X-3 are fainter ($m_{V} \sim 26.9$; $A_{V}=1.5$). The obtained CMD (Figure 5) shows while stars in the cluster would have (V$-$I) colors are between $-$0.2 to 1.7, the color values for the optical candidates are between 1.5 and 2. These values correspond to the upper limit of the color range of the stars. Using the isochrones from the PARSEC code (Bressan et al. 2012) and taking into account extinctions $E(\bv)=0.48$ mag and $E(V-I)= 0.60$ mag based on the $A_{V}$(=1.5), we derive the age of the stars as $\leq$ 30 Myr on the other hand, the age and mass of the possible optical counterparts of X-3 were estimated as (40$-$63 Myr) and $\sim 7$ M$\sun$, respectively. Determined age interval indicates that the possible candidates quite old compared to the stars in the cluster. In addition to the age difference, the position of the counterparts on the isochrones and derived spectral type (somewhat evolved F or G type supergiant) make them difficult to be member of the cluster.

\textbf{X-4 :}
As Figure 6 shows a unique possible optical counterpart is visible for X-4. This bright source is clearly observable in all images, making this source the most distinctively identified candidate. In its environment there is a small group of stars $\sim 1\farcs2$ away from the source. We obtained CMDs to estimate the age of candidate and the star group by using ($A_{V}=1$). The updated version of the code used to compute stellar tracks, leads to nonphysical mass-loss and inconsistencies between the mass and spectral type of donor star (1.4$-$3 $M_{\sun}$ indicates B6$-$B9 supergiant). These results may be due to the high extinction hence high reddening of $E(\bv)=0.31$ mag towards X-4. To cope with this issue we estimated another $A_{V}$ from the following way: The interstellar reddening values of the stars have been derived by aligning the PARSEC isochrones with the reddening line in the F555W-(F438W$-$F555W) color-magnitude diagram. The corresponding color excesses applied to the diagram F814W versus F555W$-$F814W by using the standard interstellar extinction laws (Pandey et al. 2003). The best fit $A_{V}$ and $A_{I}$ values are determined as 0.5 and 0.24 correspond to $E(\bv)=0.16$ and $E(V-I)= 0.2$ respectively. Dereddened magnitudes of the possible optical counterpart of X-4 are given in Table 5 and a created one of the CMDs is shown in Figure 7, for F814W versus F555W$-$F814W. The age and the mass of the candidate star from both CMDs were estimated based on lower extinction value as 28 Myr and 9 $M_{\sun}$, respectively. The spectral type of the possible donor of X-4 determined as A1$-$A3 supergiant. It is seen that the values estimated using the lower absorption yielded much more reasonable results. 

There is also a group of stars that is $\sim 1\farcs2$ away from X-4. The (V$-$I) colors of majority of stars in this group are in the range of $-0.5$ - $1.5$ and absolute magnitudes, $-3.5 < M_{V} <-6$. These stars may be late O type or early B stars evolving toward the supergiant phase. It is seen from CMD (Figure 7) the candidate counterpart of X-4 may belong to these nearby OB associations. Several ULXs which were found to be associated with OB stars are well studied in the literature (Abolmasov et al. 2007a, Liu et al. 2007, Grise et al. 2012). By comparing the counterpart position on the diagrams with Padova stellar models, we found that the possible optical counterpart has an age of 28 Myr from its B(F435W) and V(F555W) magnitudes after correction for $E(\bv)=0.16$ mag and has the same age from its V(F555W) and I(F814W) magnitudes after correction for $E(V-I)=0.20$. We note that almost all of the stars in the group are probably older than the candidate and located near the 25$-$75 Myr isochrones for both CMDs. However, the high probability about several ULXs ejected from nearby star cluster prompt us to keep this ULX-star group association under our attention.
 
\textbf{X-6 :}
Three possible optical counterparts were identified for this source within its error circle (Figure 8). A group of star with an extent of about 76 pc is located $\sim 1\farcs3$ to the west of X-6. Similar inconsistencies between the mass and spectral type of donor star were observed for X-6 probably due to high extinction $A_{V}=1.7$ (corresponding high reddening of $E(\bv)=0.55$). Thus, the procedure described for the source X-4 was applied. As a result, the $A_{V}=0.85$ and $A_{I}=0.41$ for the possible donor star of X-6 were determined and these extinction values yield $E(\bv)=0.28$ and $E(V-I)=0.34$. Extinction corrected magnitudes of the possible optical counterparts are given in Table 5. A CMD for possible counterparts and nearby stars is given in Figure 9, for F814W versus F555W$-$F814W. Then the age of the sources were estimated as 22 Myr, 36 Myr, and 50 Myr for src1, src2, and src3, respectively. Also the mass values were estimated as 10 $M_{\sun}$, 8 $M_{\sun}$, and 7 $M_{\sun}$, respectively. The spectral types of the possible optical counterparts are found as B1$-$A3 type supergiant for src1, G8$-$K3 supergiant for src2 and O$-$B6 supergiant for src3.

The group of stars near X-6 contains both blue and red stars. According to the CMD (Figure 9), the age of these stars are $>22$ Myr and the (V$-$I) color value is in the range of -0.5 $-$ 2 and (V$-$I) color of the candidates are 0.1 $-$ 1.7. The age and color values of the optical candidates of X-6 and the stars in the group are compatible with each other. 

\textbf{X-7 :}
This is the only ULX source identified as a transient in our ULX list. Two sources are seen in the error circle in the {\it HST}/WFC3 F814W image of X-7 (Figure 10). The fainter one, which is the center of the red circle, was not determined in other filters. Therefore, the source seen in all filters was taken as possible optical candidate of X-7. The dereddened magnitudes of candidate are given in Table 5. Due to the location of the source, the surroundings are very crowded. There is no significant star cluster structure. Therefore, a region of bright stars $2\farcs5$ away to the north of X-7 is selected. The resultant CMD obtained for F814W versus F555W$-$F814W is given in Figure 11. The age and mass of companion star are derived as 28$-$32 Myr and 9 $M_{\sun}$. The spectral type of optical candidate is determined as A3$-$F0 supergiant.
 
The majority of the stars detected near ULX X-7 are bright blue stars. The absolute magnitude values of these stars are in the range of $-8.5>M_{V}>-6$ while the magnitude of the possible optical counterpart of X-7 is $M_{V}=-5.5$. According to the obtained CMD (Figure 11), these bright stars in the region are younger than the candidate. The differences in the magnitude and age values do not support the association of the ULX with the group.

\subsection{X-ray observations}
The NGC4490/4485 pair has been observed three times with {\it Chandra} from 2000 to 2004 and four times with {\it XMM-Newton} from 2002 to 2015. Previously, the detailed spectral analyses for all ULXs (except the transient one) were carried out by Gladstone \& Roberts (2009) and Yoshida et al. (2010) using the available datasets ({\it Chandra} data (ObsID 1579, 4725, 4726) and one {\it XMM-Newton} data (ObsID 0112280201)). In the present study, we used three archival {\it XMM-Newton} datasets (taken in 2008 May (ObsID 0556300101), 2008 June (ObsID 0556300201), and 2015 June (ObsID 0762240201)) not used before. These datasets were analysed for four sources (X-2, X-3, X-4 and X-6) since X-7 was detected only by one {\it Chandra} observation (ObsID 1579). In addition, we also reanalyzed the all older data sets mentioned. 

As the standard data analysis for {\it XMM-Newton} we used {\scshape science analysis software} ({\scshape sas} version 13.05). The event files were created using {\scshape epchain} and {\scshape emchain} tasks in {\scshape sas}. Two 2008 observations were affected significantly by several background flaring which were excluded in analysis. This process reduced 2008 data exposures by about half. The source and background spectra files were obtained using {\scshape evselect} task and grouped to have a minimum of 20 counts per bin prior to fitting. 

In all three {\it XMM-Newton} observations, the source X-2 was significantly contaminated by a nearby source located $\sim12 \arcsec$ away. This contaminating source was not seen in previous {\it XMM-Newton} (ObsID 0112280201) and other two {\it Chandra} observations (ObsID 1579 and 4725). In order to minimize the contamination $5\arcsec$ radius circular regions were used to extract the spectra of X-2. No good spectral fitting was achieved for the source in 2015 observation due to its short exposure and small aperture around the source. On the other hand, ULX sources X-3 and X-4 were located close to each other (with an $18\arcsec$ separation). Therefore, circles with $9\arcsec$ radius were used for source apertures. This aperture value is in line with the aperture that Gladstone \& Roberts (2009) used for the ULXs (in NGC 4490/4485). Besides, $15\arcsec$ radius circle regions were used to obtain source and background spectra of X-6. 

We performed spectral fittings using {\scshape xspec} software (version 12.9.1). EPIC pn and MOS spectra were fitted simultaneously in 0.3$-$10 keV energy band and a constant scaling factor was introduced in order to consider the cross-calibration differences between the instruments. The unabsorbed flux and luminosity values were calculated in the 0.5$-$8 keV energy band to compare with previous results. The obtained best fitting model parameters for power law (PL) and disk blackbody (DISKBB) for ULXs are given in Table 6. The best-fitting parameters to the models are in the range $\Gamma \sim 1.8-2.4$ for the PL model, and $T_{\mathrm{in}}\sim 1.1-1.8$ keV for the DISKBB model. For each source the range of obtained spectral properties is quite similar with Gladstone \& Roberts (2009) and Yoshida et al. (2010). Therefore relevant figures are not given.

We also obtained the light curves for four ULXs using all {\it Chandra} and {\it XMM-Newton} observations to check their long-term time variability as given in Figure 12. For this we calculated unabsorbed flux values (in the 0.5$-$8 keV band) by fitting PL model.

\section{Discussion and Conclusions}

Optical properties of seven ULXs in the galaxy pair NGC 4490/4485 were studied for the first time using {\it HST} archival data. Possible optical counterparts were found only for five of them (X-2, X-3, X-4, X-6 and X-7). For two ULXs (X-4 and X-7), we identified a single candidate, remaining three (X2, X3 and X-6) have more than one counterparts. Furthermore, we have studied the environments around these five ULXs in the NGC4490/4485 galaxies. The correlation between ULX source positions and stellar clusters has been already noticed in the literature (see Kaaret et al. 2004; Abolmasov et al. 2007a; Grise et al. 2012; Avdan et al 2016a). Several authors have suggested that some ULXs might have their clusters already dispersed due to their ages (Liu et al. 2007; Poutanen et al. 2013). 

We also use the available high quality $\mathrm{H}\alpha$ image from {\it HST}/WFC3 archive in order to probe the current star formation regions and check for possible ULX nebulae which might supply important information about the intrinsic photon luminosity and wind/jet power (Feng \& Soria 2011). The positions of ULXs on F657N image show that only a nebular structure appeared around the source X-7. ULX X-7 appears almost at the south of the nebular structure whose diameter is $\sim$200 pc ($\sim 5\arcsec$). We have examined the north (more bright) and south part of the nebula near the ULX X-7 (Figure 13). The aperture photometry was performed to find absolute $\mathrm{H}\alpha$ flux with apphot package in IRAF for the F657N image from {\it HST}/WFC3/UVIS (dataset no: ICNK19010). We have measured the total flux of both parts of nebula with stars in a circle aperture with radius 45 pixel ($1.8 \arcsec$). For the background calculation an annulus (with inner radius of 80 pixel and outer radius of 150 pixel) around the circle region was chosen and the contribution of stars (with 2 pixel aperture) were removed to obtain continuum subtracted nebular flux. The calculated continuum subtracted flux values from the northern and southern parts of nebula are $\sim$ $5.20\times10^{-16}$ erg cm$^{-2}$ s$^{-1}$ and $2.73\times10^{-16}$ erg cm$^{-2}$ s$^{-1}$, respectively. After the reddening corrections (for A$_{V}$=2.1), the flux value is $2.53\times10^{-15}$ erg cm$^{-2}$ s$^{-1}$ for the northern nebula and $1.33\times10^{-15}$ erg cm$^{-2}$ s$^{-1}$ for the southern nebula. These reddening corrected fluxes correspond luminosities $\sim$ $1.84\times10^{37}$ erg s$^{-1}$ and $9.65\times10^{36}$ erg s$^{-1}$ at distance of 7.8 Mpc. Statistical errors of the flux values are about or less than 2\%. It needs to be noted that background level can vary considerably depending on the selected region. This could ultimately lead to the uncertainty of measuring the nebular flux at the level of 30\%.
In case of X-7, we don't observe a brightening of the nebula near the ULX position as observed in the nebulae around the NGC 6946 ULX-1 (Abolmasov et al. 2008), HolmbergII X-1 (Lehmann et al. 2005) and NGC 5408 X-1 (Grise at al. 2012). Note that also these sources and their nebulae are very bright and compact (a few tens of parsec). On the other hand, the ULX nebulae of IC 342 X-1 and HolmbergIX X-1 are very large size (up to 500 pc). The common feature of these nebulae is that they have high luminosity (about 10$^{38}$ in $\mathrm{H}\alpha$ and more than about 10$^{39}$ erg s$^{-1}$ bolometric; Abolmasov et al. 2007). In case of the X-7 nebula has a size about 200 pc, but its $\mathrm{H}\alpha$ line luminosity is less by a factor of 3$-$10 (compared to the sources listed in Table 2 in Abolmasov et al.2007). We think that UV luminosity of young nearby star cluster is enough to photoionize the gas of this nebula, but the contribution from the ULX source is not clear. We also keep in mind that if there is time dependent illumination and ionization (X-7 is transient source), it may not show characteristics of the other brighter $\mathrm{H}\alpha$ nebulae around ULXs. This matter requires more study in the future work. Particularly, we need nebula spectrum to model with CLOUDY spectral synthesis code.

On the other hand X-ray data analyses provide us with good fits for PL and DISKBB models to explain their spectra. The obtained long-term light curves (Figure 12) suggest that the flux of the sources were changed within less than a factor of three over the observation duration. Considering the highest luminosity values for each source in Figure 12 and assuming, in that epoch the sources emit at the Eddington limit, the masses of the compact objects can be estimated. With this approach, the masses of the compact objects in ULXs might be range of (30$-$50) $M_{\sun}$. Additionally, another mass value was calculated for the compact object by considering the normalization parameter obtained from the DISKBB model in {\scshape xspec} for four ULXs(X-2, X3, X-4 and X-6). The apparent inner disk radius values ($r_{\mathrm{in}}$) were corrected using the equation $R_{\mathrm{in}}=\xi \cdot \kappa^{2} \cdot r_{\mathrm{in}}$. Here, correction factor $\xi = 0.412$, spectral hardening factor $\kappa = 1.7$ and $R_{\mathrm{in}}$ is the true inner disk radius (Shimura \& Takahara 1995; Kubota et al. 1998). Assuming a disk inclination of 60 degree and using the relation between mass and inner disk radius (Makishima et al. 2000), we calculated the masses of the compact objects in the range between 10 to 15 $M_{\sun}$. These values are compatible with the results presented in previous studies (Gladstone et al. 2009; Yoshida et al. 2010) and consistent with stellar mass black holes. 

Power Density Spectra (PDS) of the ULXs have been obtained using Fast Fourier Transform on the consideration of the compact object as a neutron star. We used the mentioned three XMM-Newton datasets for these analyses. PDS of these sources do not show a prominent peak at any frequency value. This may indicate that there is probably no neutron star involved as a compact object.

In order to explore further, theoretical values derived from binary evolution models including the effects of irradiation (Patruno \& Zampieri 2008, 2010) were compared with the observed $M_{V}$ and $(B-V)_{0}$ values to constrain the properties of the mentioned donor stars. Taking into account the errors on V and B, the data for X-2, X-6 (only for their possible counterparts src1), and X-7 are best fitted with a donor of 10-15 $M_{\sun}$ during hydrogen shell burning given a black hole mass of 10 $M_{\sun}$, but in the case of a black hole of 100 $M_{\sun}$ the model shows that a 10 $M_{\sun}$ donor is compatible with the observed photometric properties of X-4 and X-6 (for src1) in the H-shell burning phase. Also, if the black hole mass is 20 $M_{\sun}$  (Patruno \& Zampieri 2010) we could restrict the candidates of ULX X-2, X-4 and X-6 (for src1) to be less than 20 $M_{\sun}$.

The high $F_{\mathrm{X}}/F_{\mathrm{opt}}$ ratio values ($\geq$100) indicate that ULXs have fainter possible optical counterparts therefore could be distinguished from Active Galactic Nuclei (AGN) (Avdan et al. 2016b ). Thus, we calculated the ratio of $F_{\mathrm{X}}/F_{\mathrm{opt}}$ for all ULXs. For this calculation, the closest date observations between {\it XMM-Newton} (ObsID 0762240201) and {\it HST} (ObsID JC9V64010 for X-2 and  ICDM42060 for other ULXs) were used for the lack of simultaneous X-ray and optical observations for these sources. $F_{\mathrm{X}} / F_{\mathrm{opt}}$ ratio values for all ULXs are given in Table 4 and Table 5, where $F_{\mathrm{X}}$ is the observed flux between 2$-$10 keV and $F_{\mathrm{opt}}$ is the optical flux in V band. The ratio values found are within the range of minimum (500) and maximum (7800) values. These values are significantly higher than those found for AGN ($0.1 \leq F_{\mathrm{X}} / F_{\mathrm{opt}} \leq 10$, Aird et al. 2010) and also for the extreme cases of heavily obscured AGNs, $\sim$ 100, (Della Ceca et al. 2015). The estimated average $F_{\mathrm{X}} / F_{\mathrm{opt}}$ ratio values for well-studied ULXs are $\sim$ 1600 (Avdan et al. 2016b and references therein). Even though only two of the ULXs (X-3 and X-6) have higher values than this average ratio, the remaining sources still have ratio values higher than $ > 100$.

An alternative definition of X-ray to optical flux ratio has been proposed as $\xi$ to distinguish between low mass X-ray binaries and high mass X-ray binaries. This ratio is defined as $\xi = B_{0} + 2.5$log$F_{\mathrm{X}}$, where $F_{\mathrm{X}}$ is the observed X-ray flux density in 2-10 keV units of $\mu$Jy and $B_{0}$ is the dereddened B magnitude (van Paradijs \& McClintock, 1995). To determine $\xi$, the flux densities were calculated using the closest date {\it XMM-Newton} observations available in archives. $F_{\mathrm{X}}$ values obtained for X-3, X-4 and X-6 are 0.066 $\mu$Jy, 0.051 $\mu$Jy and 0.047, respectively. Thus, the $\xi$ values were found to be $\sim$20$-$23 for X-3, $\sim$20 for X-4 and $\sim$20$-$22 for X-6. This value was not calculated for X-7 because the source was not detected in any {\it XMM-Newton} data. The color ratios are given in the range of (12$-$18) for high mass X-ray binaries and (21$-$22) for low mass X-ray binaries (van Paradijs \& McClintock, 1995). For the given $\xi$ values, the X-3 and X-4 might be identified as low mass X-ray binaries like most of the ULXs studied so far (Tao et al. 2011; Kaaret et al. 2017). As discussed in Tao et al. (2011) this does not mean that ULXs actually have low mass companions, instead may suggest that the optical spectra of ULXs are dominated by emission from the disk but not from the companion star. 

As discussed in the related studies, the optical emission from the ULXs could be arise from the accretion disk or the donor star or both. Assuming that the light from the optical counterpart is dominant, then the optical variability should be $\leq$ 0.1 mag and consistent with a single spectral type in all observations. On the other hand, if the optical emission is dominated by the accretion disk, optical emission may exhibit significant variability. Unfortunately, there are not sufficient optical data to discuss these conditions for the sources examined in this study.
 
In order to improve our understanding of the nature of these sources, more sensitive photometric and spectroscopic observations should be performed.

\acknowledgments

The authors thank the anonymous referee for helpful suggestions and comments that improved the manuscript. We also would like to thank M. Emin Ozel for his useful comments. This research was  supported the Scientific and Technological Research Council of Turkey (TUBITAK) through project number 117F115 and by Cukurova University Research Fund through project number FDK-2014-1998. The research was also supported by the Russian RFBR grants 16-32-00210,  16-02-00567,  and  the  Russian  Science  Foundation  grant  N 14-50-00043  for  observations  and  data reduction.

\clearpage

\begin{table}
\centering
\caption{The Log of {\it HST}/WFC3 and ACS observations of ULXs}
\begin{tabular}{cccccc}
\hline\hline
  
Filter & ObsID & Date & Exp. \\
\hline
& X-2  & & \\
ACS/F435W  & J9H807010 & 2005-11-19 & 1116 \\
ACS/F606W  & J9H807020 & 2005-11-19 & 1116 \\
WFC3/F275W & ICDM51030 & 2014-01-12 & 2361 \\
WFC3/F336W & ICDM51040 & 2014-01-12 & 1119 \\
WFC3/F814W & ICDM51050 & 2014-01-12 & 965  \\
ACS/F606W  & JC9V64010 & 2014-08-05 & 1000 \\
ACS/F814W  & JC9V64020 & 2014-08-05 & 1000 \\
\hline
& X-3, X4, X-6, X-7  & & \\
WFC3/F275W & ICDM42030 & 2013-10-30 & 2361 \\
WFC3/F336W & ICDM42040 & 2013-10-30 & 1107 \\
WFC3/F438W & ICDM42050 & 2013-10-30 & 953  \\
WFC3/F555W & ICDM42060 & 2013-10-30 & 1131 \\
WFC3/F814W & ICDM42070 & 2013-10-30 & 977  \\
\hline
\end{tabular}
\end{table}

\begin{table}
\centering
\caption{Coordinates and their uncertainties at 90\% confidence level of the two X-ray/optical reference sources and ULXs}
\begin{tabular}{cccccc}
\hline\hline
  
Object & RA & Dec & Position uncertainty in arcsec  \\
\hline
\multicolumn{4}{c}{Chandra coordinates of the reference sources}\\
s1 &  12:30:49.507 & +41:40:56.81 & 0.12  \\
s2 &  12:30:32.463 & +41:43:29.46 & 0.54  \\
\hline
\multicolumn{4}{c}{Chandra coordinates of the ULXs}\\
X2 &  12:30:30.493 & +41:41:42.32 & 0.07  \\
X3 &  12:30:30.754 & +41:39:11.72 & 0.08  \\
X4 &  12:30:32.204 & +41:39:18.24 & 0.07  \\
X5 &  12:30:36.253 & +41:38:38.02 & 0.07  \\
X6 &  12:30:43.180 & +41:38:18.72 & 0.07  \\
X7 &  12:30:38.437 & +41:38:31.91 & 0.10  \\
\hline
\multicolumn{4}{c}{SDSS coordinates of the reference sources}\\
s1 &  12:30:49.477 & +41:40:56.90 & 0.08  \\
s2 &  12:30:32.440 & +41:43:29.60 & 0.08  \\
\hline
\multicolumn{4}{c}{Corrected ULXs coordinates on SDSS image}\\
X2 &  12:30:30.462 & +41:41:42.41 & 0.20  \\
X3 &  12:30:30.723 & +41:39:11.81 & 0.21  \\
X4 &  12:30:32.173 & +41:39:18.33 & 0.20  \\
X5 &  12:30:36.222 & +41:38:38.11 & 0.20  \\
X6 &  12:30:43.149 & +41:38:18.81 & 0.19  \\
X7 &  12:30:38.406 & +41:38:32.00 & 0.21  \\
\hline
\multicolumn{4}{c}{Corrected ULX X2 coordinates on J9H807020 image}\\
X2 &  12:30:30.526 & +41:41:43.10 & 0.21  \\
\hline
\multicolumn{4}{c}{Corrected ULXs coordinates on ICDM42060 image}\\
X3 & 12:30:30.716 & +41:39:11.91 & 0.22   \\
X4 & 12:30:32.166 & +41:39:18.43 & 0.21   \\
X5 & 12:30:36.215 & +41:38:38.21 & 0.20   \\
X6 & 12:30:43.142 & +41:38:18.91 & 0.19   \\
X7 & 12:30:38.399 & +41:38:32.10 & 0.21   \\
\hline
\end{tabular}
\end{table}

\begin{table}
\centering
\caption{The calculated extinction values ($A_{V}$) and the distance between the nebulae and the ULXs}
\begin{tabular}{cccccc}
\hline\hline
Source & Distance & $A_{V}$ \\
& $\arcsec$ & \\
\hline
X-2 & 14.8 & $1.0\pm0.3$ \\
X-3 & 7.9  & $1.5\pm0.2$\\
X-4 & 22.0   & $1.0\pm0.1$\\
X-6 & 28.4 & $1.7\pm0.2$ \\
X-7 & 25.5 & $2.1\pm0.2$ \\
\hline
\end{tabular}
\end{table}

\begin{table}
\centering
\caption{The calculated {\it HST}/ACS and WFC3 dereddened magnitude values of the possible optical counterpart of X-2 (the values are given chronologically, stars (*) in the table indicate that this source is not seen)}
\begin{tabular}{cccccc}
\hline\hline
&  \multicolumn{2}{c}{X-2} & \\
\hline
& src1 & src2 & src3 \\
\hline
ACS/F435W & $24.85\pm 0.18$ & $*$ & $*$ \\
ACS/F606W & $24.87\pm 0.11$ & $25.52\pm 0.14$ & $25.71\pm 0.23$ \\
WFC3/F814W & $24.52\pm 0.14$ & $*$ & $*$ \\
ACS/F606W & $24.85\pm 0.10$ & $25.57\pm 0.18$ & $25.82\pm 0.33$ \\
ACS/F814W & $24.85\pm 0.14$ & $24.78\pm 0.12$ & $25.63\pm 0.27$ \\
$(B-V)_{0}$ & -0.02  &  &   \\
$(V-I)_{0}$ & 0  & 0.79 & 0.19  \\
$M_{V}$ & -4.60 & -3.96 & -3.71 \\
F$_{X}$/F$_{opt}$\tablenotemark{*} & 1200 & 2400 & 3000 \\ 
\hline
\end{tabular}
\tablenotetext{*}{These ratio were calculated in the range of wavelength (4584$-$6209) \AA\ for F$_{opt}$ and energy band 2$-$10 keV for F$_{X}$.}
\end{table}

\begin{table*}
\centering
\caption{The calculated {\it HST}/WFC3 dereddened magnitude values of the possible optical counterparts of X-3, X-4, X-6 and X-7 (Stars (*) in the table indicate that the source is not seen. F$_{\mathrm{X}}$/F$_{\mathrm{opt}}$ of X-7 is not presented since the source was not detected in {\it XMM-Newton} data.)}
\begin{tabular}{cccccccc}
\hline\hline
& \multicolumn{2}{c}{X-3} & {X-4} & \multicolumn{3}{c}{X-6} & {X-7}  \\
\hline
& src1 & src2 & & src1 & src2 & src3 & \\
\hline
WFC3/F275W & $24.68\pm 1.15$ & $*$ & $22.25\pm 0.04$ & $22.44\pm 0.08$ & $*$ & $ 23.78\pm 0.23$ & $*$ \\
WFC3/F336W & $*$ & $24.49\pm 0.69$ & $22.29\pm 0.04$ & $22.54\pm 0.12$ & $*$ & $24.02\pm 0.24$ & $ 23.43\pm 0.77$ \\
WFC3/F438W & $25.67\pm 0.56$ & $25.18\pm 0.24$ & $23.74\pm 0.05$ & $24.15\pm 0.11$ & $26.67\pm 0.86$ &$25.64\pm 0.34$ & $ 24.07\pm 0.37$  \\
WFC3/F555W & $25.40\pm 0.19$ & $25.66\pm 0.23$ & $23.74\pm 0.03$ & $24.07\pm 0.05$ & $25.08\pm 0.10$ & $25.19\pm 0.11$ & $23.88\pm 0.12 $ \\
WFC3/F814W & $24.25\pm 0.10$ & $24.81\pm 0.17$ & $23.62\pm 0.05$ & $24.46\pm 0.12$ & $23.83\pm 0.08$ & $25.32\pm 0.25$ & $23.91\pm 0.14$  \\
$(B-V)_{0}$ &  0.27  & -0.48  & 0.001 & 0.08 & 1.59 & 0.36 & 0.19  \\
$M_{V}$ & -4.06  & -3.80 & -5.72 & -5.39 & -4.38 & -4.27 &-5.58  \\
F$_{\mathrm{X}}$/F$_{\mathrm{opt}}$\tablenotemark{*} & 7800 & 9890 & 500 & 775 & 1950 & 2160 & $*$ \\
\hline
\end{tabular}
\tablenotetext{*}{These ratio were calculated in the range of wavelength (4584$-$6209) \AA\ for F$_{opt}$ and energy band 2$-$10 keV for F$_{X}$.}
\end{table*}

\begin{table*}
\centering
\caption{The best fitting model parameters for the ULXs in NGC 4490/4485}
\begin{tabular}{ccccccccc}
\hline\hline
ObsID & Model & $N_{\mathrm{H}}$ & $\Gamma$ & $kT_{\mathrm{in}}$ & $\chi^{2}$/dof & $L_{\mathrm{X}}$\tablenotemark{*}  \\
  & & $10^{22}$ cm$^{-2}$ & & keV &  & $10^{39}$ erg s$^{-1}$ \\
\hline
\multicolumn{7}{c}{X-2}\\
& & & & & & \\
        
0556300101 & PL & $0.24^{+0.02}_{-0.02}$ & $2.04^{+0.08}_{-0.08}$ & $-$ & 47.36/63 & $4.00^{+0.26}_{-0.26}$ \\
    & DISKBB & $0.03^{+0.02}_{-0.02}$ & $-$ & $1.28^{+0.08}_{-0.08}$ & 60.74/63 & $2.84^{+0.19}_{-0.18}$ \\

0556300201 & PL & $0.27^{+0.04}_{-0.03}$ & $2.23^{+0.12}_{-0.12}$ & $-$ & 34.00/32 & $2.76^{+0.25}_{-0.25}$ \\
    & DISKBB & $0.04^{+0.03}_{-0.03}$ & $-$ & $1.13^{+0.11}_{-0.09}$ & 34.09/32 & $1.87^{+0.16}_{-0.16}$ \\

\hline
    \multicolumn{7}{c}{X-3}\\
    & & & & & & \\

0556300101 & PL & $0.81^{+0.05}_{-0.05}$ & $1.80^{+0.07}_{-0.07}$ & $-$ & 114.17/106 & $5.21^{+0.28}_{-0.28}$ \\
    & DISKBB & $0.45^{+0.05}_{-0.04}$ & $-$ & $1.86^{+0.12}_{-0.11}$ & 117.35/106 & $3.53^{+0.18}_{-0.19}$ \\

0556300201 & PL & $1.24^{+0.09}_{-0.09}$ & $2.17^{+0.09}_{-0.09}$ & $-$ & 78.98/84 & $6.05^{+0.38}_{-0.45}$ \\
    & DISKBB & $0.71^{+0.09}_{-0.09}$ & $-$ & $1.51^{+0.12}_{-0.11}$ & 86.80/84 & $3.47^{+0.25}_{-0.25}$ \\
    
0762240201 & PL  & $0.80^{+0.06}_{-0.06}$ & $2.14^{+0.09}_{-0.09}$ & $-$ & 47.04/57 & $6.42^{+0.45}_{-0.39}$ \\
    & DISKBB & $0.41^{+0.06}_{-0.05}$ & $-$ & $1.40^{+0.10}_{-0.09}$ & 43.97/57 & $4.06^{+0.27}_{-0.27}$ \\
       
    \hline
    \multicolumn{7}{c}{X-4}\\
    & & & & & & \\

0556300101 & PL & $0.51^{+0.04}_{-0.04}$ & $2.05^{+0.08}_{-0.08}$ & $-$ & 79.73/79 & $3.28^{+0.19}_{-0.20}$ \\
    & DISKBB & $0.22^{+0.03}_{-0.03}$ & $-$ & $1.41^{+0.09}_{-0.08}$ & 96.77/79 & $2.19^{+0.13}_{-0.13}$ \\

0556300201 & PL & $0.74^{+0.06}_{-0.05}$ & $2.20^{+0.09}_{-0.09}$ & $-$ & 96.24/94 & $4.90^{+0.34}_{-0.32}$ \\
    & DISKBB &  $0.36^{+0.05}_{-0.05}$ & $-$ & $1.29^{+0.10}_{-0.10}$ & 126.23/94 & $2.95^{+0.20}_{-0.21}$ \\

0762240201 & PL  & $0.70^{+0.06}_{-0.05}$ & $2.22^{+0.10}_{-0.10}$ & $-$ & 53.86/49 & $5.31^{+0.38}_{-0.38}$ \\
    & DISKBB & $0.35^{+0.05}_{-0.05}$ & $-$ & $1.23^{+0.09}_{-0.08}$ & 44.26/49 & $3.29^{+0.24}_{-0.22}$ \\
& & & & & & \\
\hline

    \multicolumn{7}{c}{X-6}\\
    & & & & & & \\

0556300101 & PL & $0.49^{+0.04}_{-0.05}$ & $2.05^{+0.09}_{-0.09}$ & $-$ & 92.77/82 & $3.82^{+0.38}_{-0.38}$ \\
    & DISKBB & $0.24^{+0.04}_{-0.04}$ & $-$ & $1.23^{+0.08}_{-0.08}$ & 75.28/82 & $2.54^{+0.25}_{-0.25}$ \\

0556300201 & PL & $0.64^{+0.04}_{-0.04}$ & $2.27^{+0.07}_{-0.07}$ & $-$ & 120.74/130 & $5.52^{+0.33}_{-0.29}$ \\
    & DISKBB &  $0.32^{+0.03}_{-0.03}$ & $-$ & $1.18^{+0.06}_{-0.06}$ & 104.80/130 & $3.46^{+0.18}_{-0.19}$ \\

0762240201 & PL  & $0.63^{+0.05}_{-0.05}$ & $2.35^{+0.09}_{-0.09}$ & $-$ & 60.78/61 & $4.85^{+0.31}_{-0.32}$ \\
    & DISKBB & $0.30^{+0.04}_{-0.04}$ & $-$ & $1.16^{+0.09}_{-0.08}$ & 66.53/61 & $2.96^{+0.17}_{-0.22}$ \\
& & & & & & \\
\hline
\end{tabular}
\tablenotetext{*}{The luminosities were calculated in the 0.5$-$8 keV energy band.}
\end{table*}

\clearpage

\begin{figure}
\label{Fig2}
\begin{center}
\includegraphics[scale=0.38]{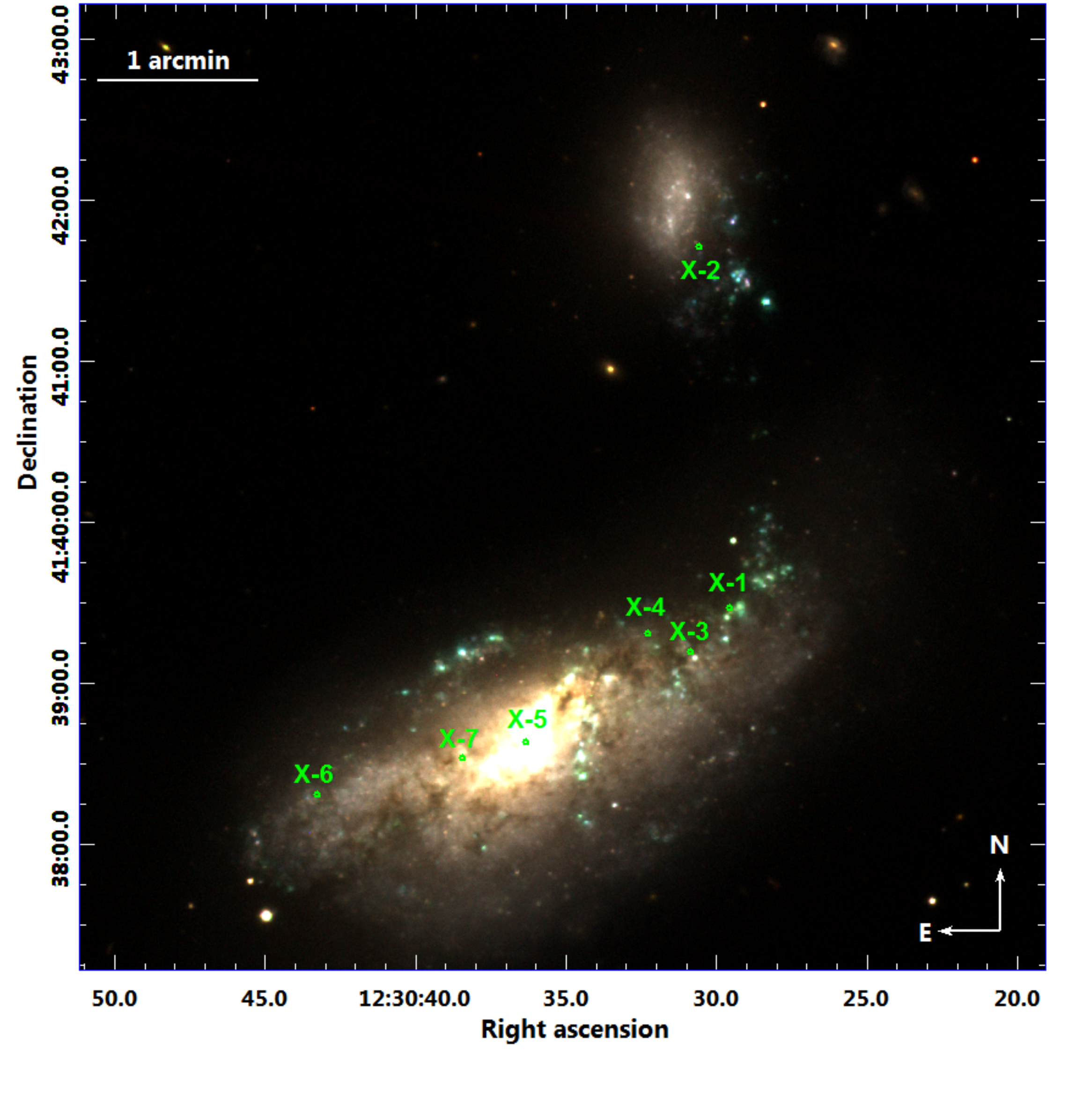}
\caption{The Sloan Digital Sky Surveys (SDSS) RGB image of NGC 4490/4485 where seven ULXs are marked on.}
\end{center}
\end{figure}

\begin{figure*}
\label{Fig2}
\begin{center}
\includegraphics[scale=0.270]{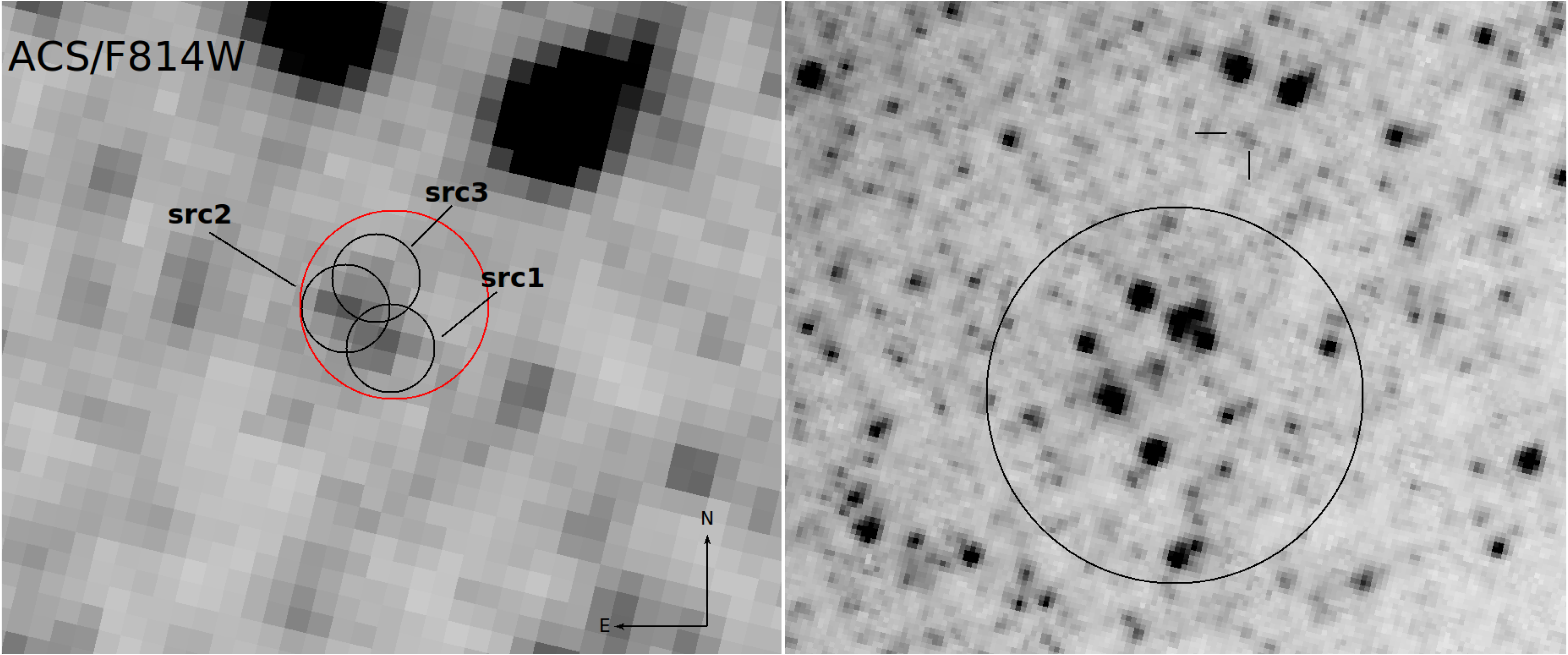}
\caption{The {\it HST}/ACS F814W (2014-08-05) images of X-2 (the zoomed image has a size of $1\farcs7\times 1\farcs2$) on the left and nearby group of star (the zoomed image has a size of $7\farcs4\times 6\farcs3$) on the right panel. The red circle represents the corrected position of X-2 with an accuracy of $0\farcs21$ radius. Three possible counterparts (black circles, the aperture of PSF fitting is $0\farcs12$ radius) are found within the error radius. In the right panel, the black circle has a radius of $1\farcs7$ and shows the group of stars. The ULX is located $\sim 2\arcsec$ away from center of the star group.}
\end{center}
\end{figure*}

\begin{figure*}
\label{Fig2}
\begin{center}
\includegraphics[scale=0.41]{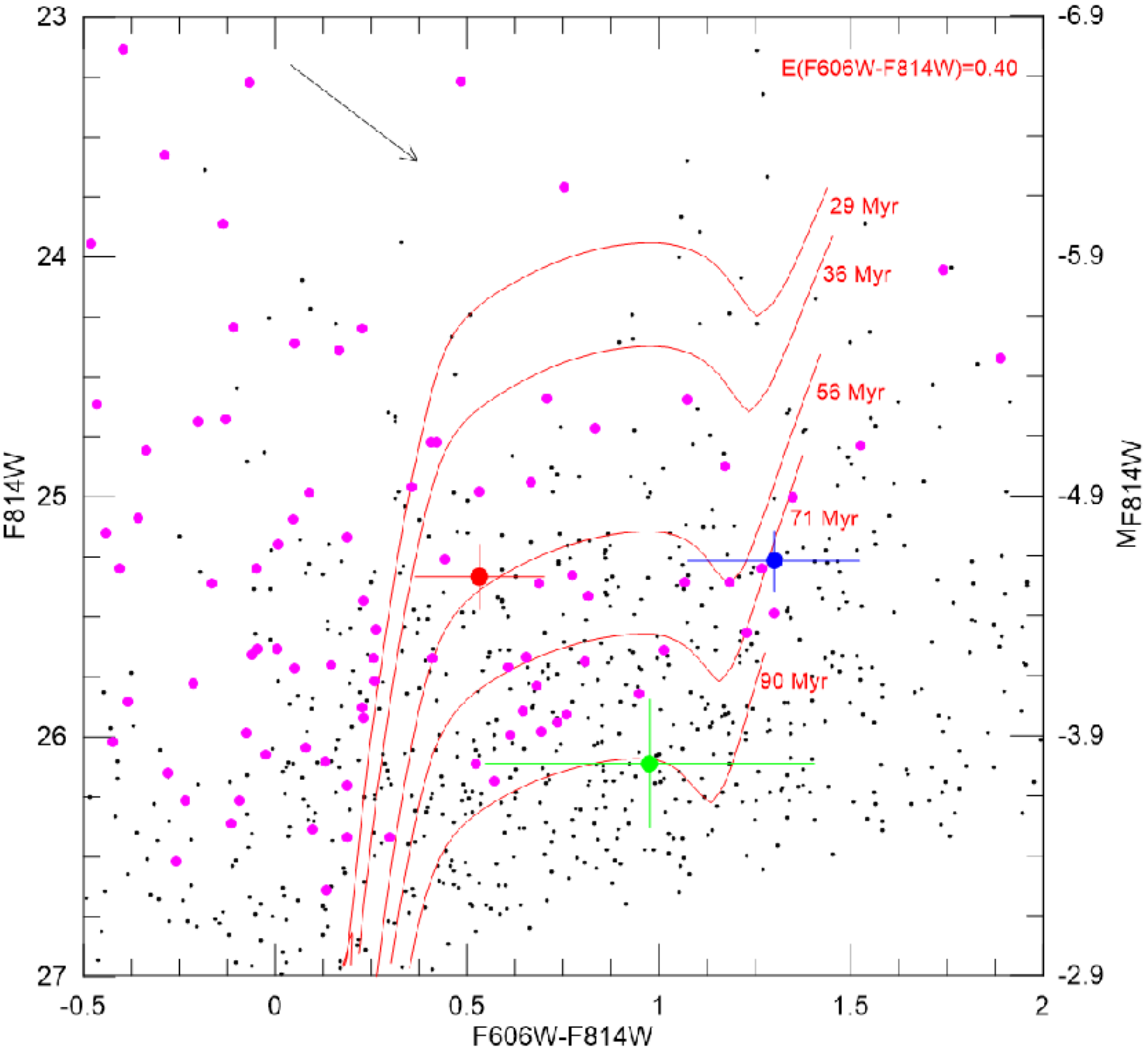}
\caption{The CMD for the possible counterparts, group of stars and field stars around X-2. The red, blue and green circles represent src1, src2 and src3, respectively. The black dots represent the field stars within the $2\arcsec$ region around the source and the magenta dots show the stars in the group. The isochrones have been corrected for extinction of $A_{V}=1.0$ mag. The black arrow shows the reddening line.}
\end{center}
\end{figure*}

\begin{figure*}
\label{Fig2}
\begin{center}
\includegraphics[scale=0.27]{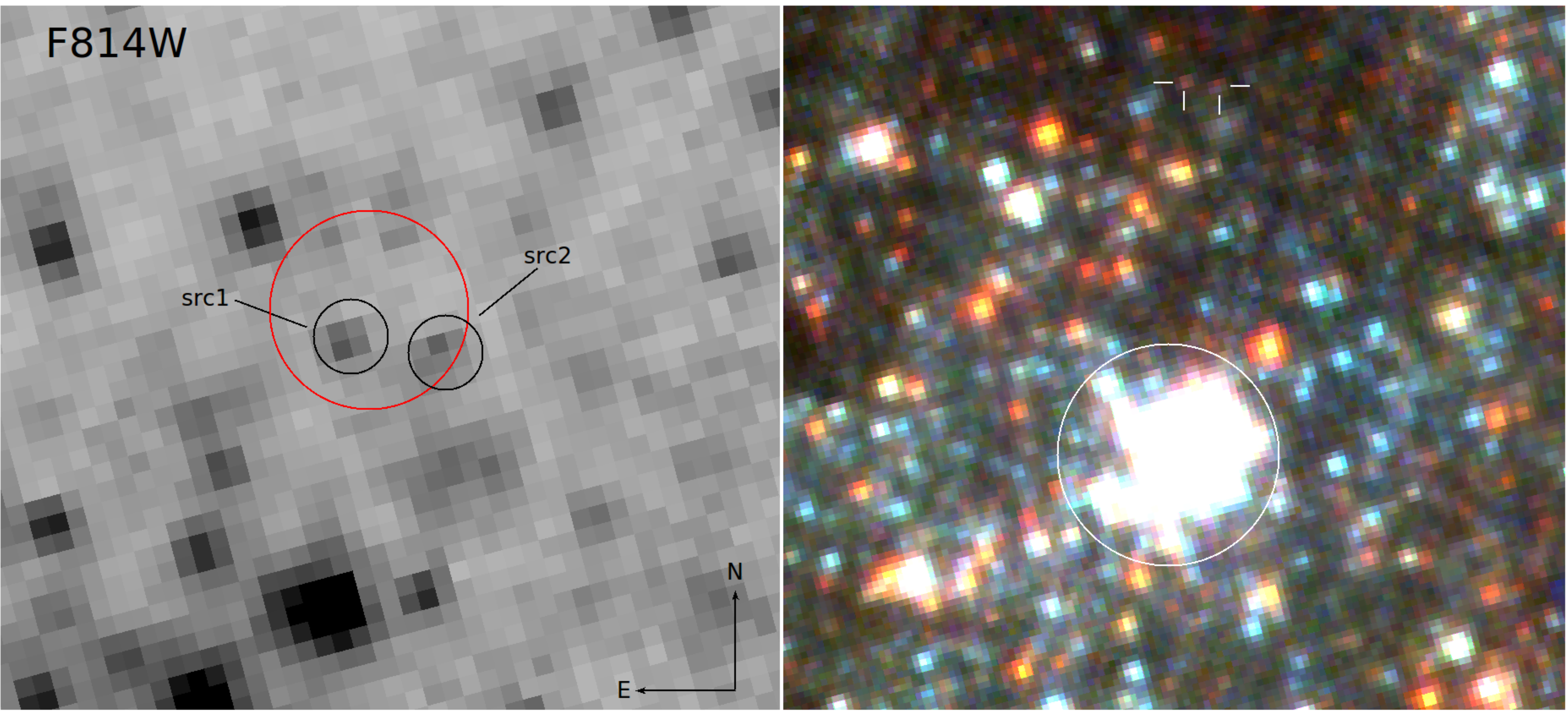}
\caption{The {\it HST}/WFC3 F814W image of X-3 (left) and the true color {\it HST} image of the environment near X-3 (Red: F814W, Green: F555W, Blue: F438W) (right). In the left panel, the red circle represents the corrected position of X-3 with an accuracy of 0\farcs22 radius. Two counterparts (black circles, the aperture of PSF fitting is 0\farcs12 radius) are found within the error circle. The zoomed image has a size of $\sim 1\farcs6 \times 1\farcs4$. In the right panel, white solid circle (1\farcs5 diameter) represents super-star cluster 2\farcs3 away from X-3. The optical candidates are marked with white bars. The RGB image has a size of $\sim 5 \arcsec \times 4 \arcsec$. }
\end{center}
\end{figure*}

\begin{figure*}
\label{Fig1}
\begin{center}
\includegraphics[angle=0,scale=0.43]{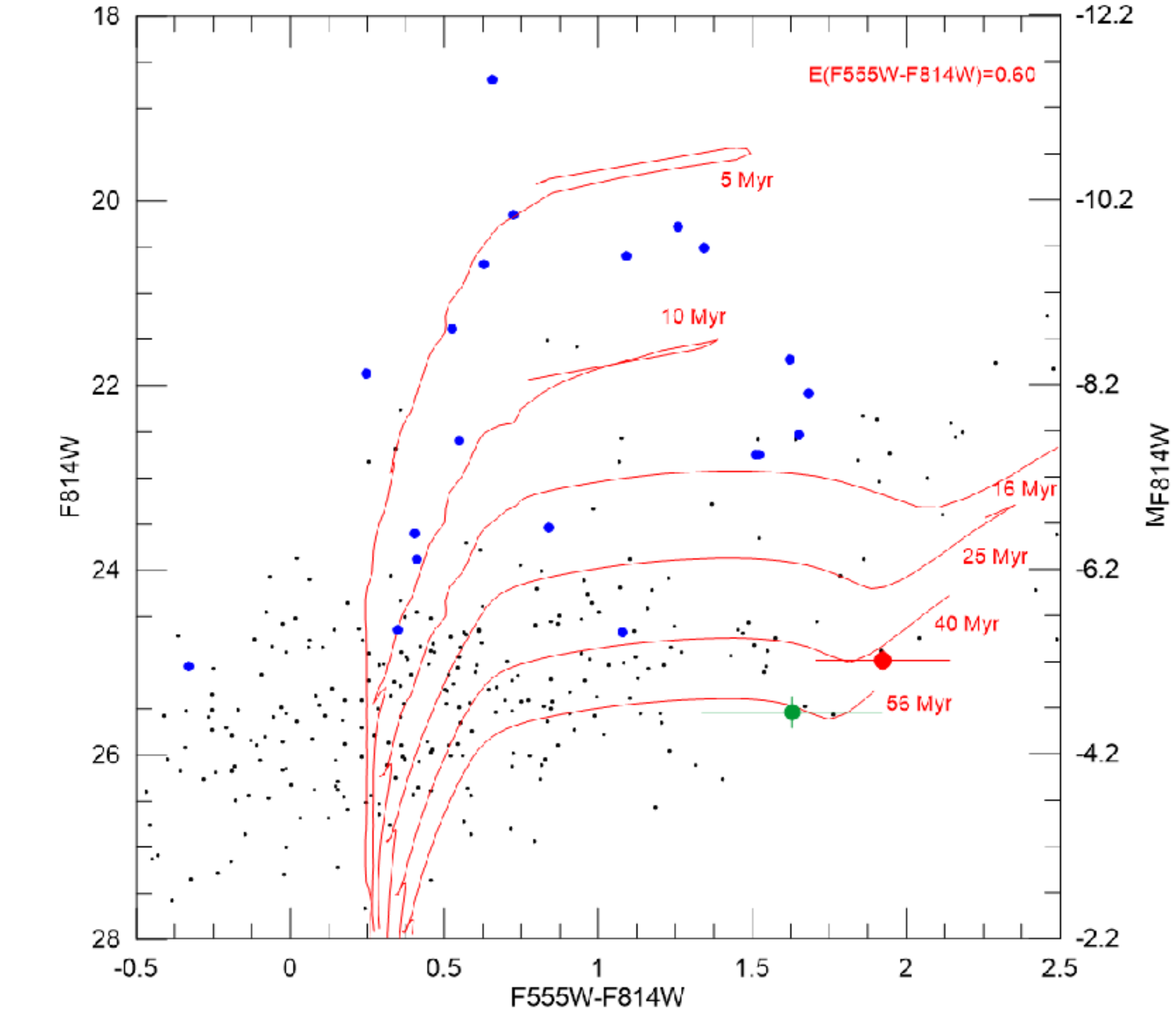}
\caption{The CMD for the possible counterparts, super star cluster and field stars around X-3. The red and green circles represent the possible optical counterparts of X-3. The black and blue dots represent the field stars within the 2\arcsec region around the X-3 and the cluster stars, respectively. X-3 is located 2\farcs3 away from center of cluster. The isochrones have been corrected for extinction of $A_{V}=1.5$ mag and the black arrow shows the reddening line. }
\end{center}
\end{figure*}

\begin{figure*}
\label{Fig2}
\begin{center}
\includegraphics[scale=0.27]{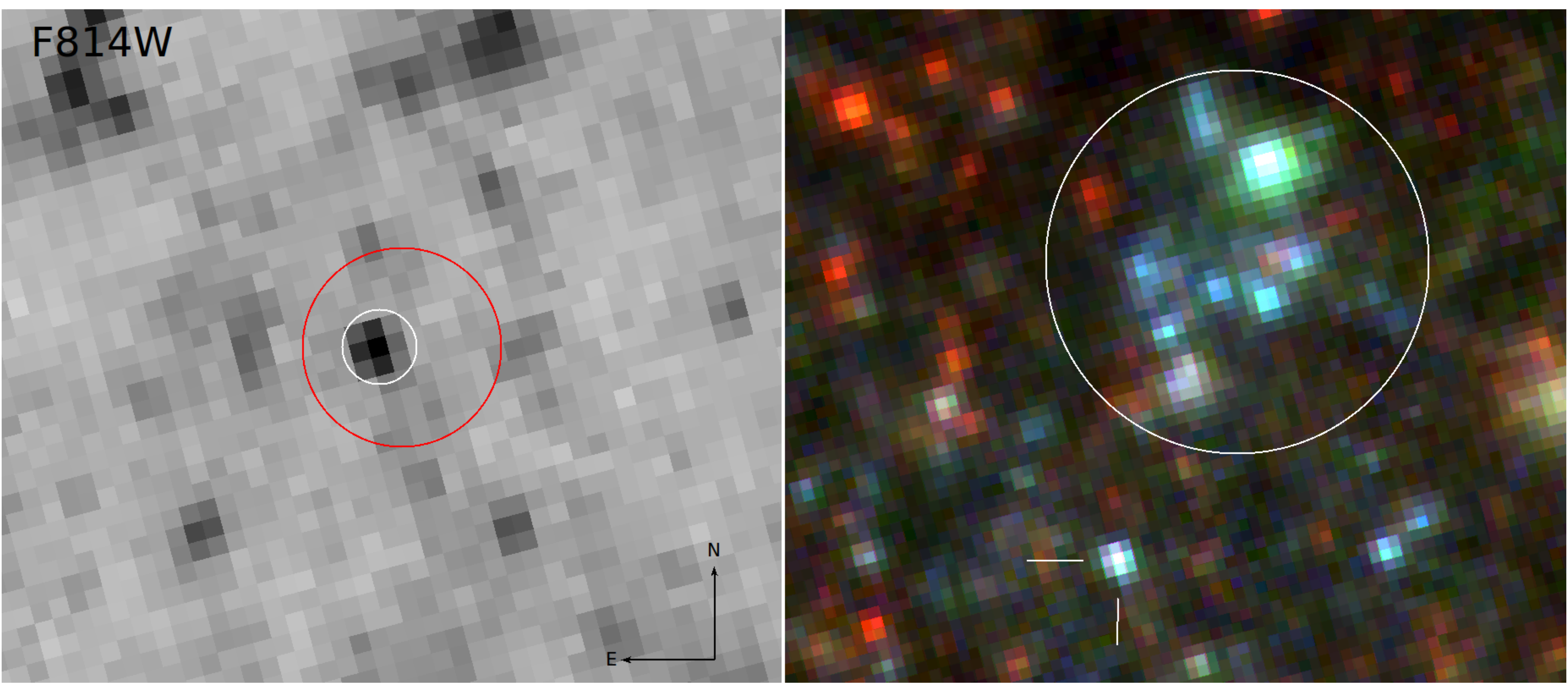}
\caption{The {\it HST}/WFC3 F814W image of X-4 (left) and the true color {\it HST} image of the star group near X-4 (Red: F814W, Green: F555W, Blue: F438W) (right). In the left panel, the red circle represent the corrected position of X-4 with an accuracy of 0\farcs21 radius. A single counterpart (white circle, the aperture of PSF fitting is 0\farcs12 radius) is found within the error circle. The zoomed image has a size of $\sim 1.6\arcsec \times 1.3\arcsec$. In the right panel, white circle (1\farcs4 diameter) represents the group of stars $\sim 1\farcs2$ away from center X-4. The optical candidate is marked with white bars. The RGB image has a size of $\sim 5\arcsec \times 3.6\arcsec$.}
\end{center}
\end{figure*}

\begin{figure*}
\label{Fig1}
\begin{center}
\includegraphics[angle=0,scale=0.40]{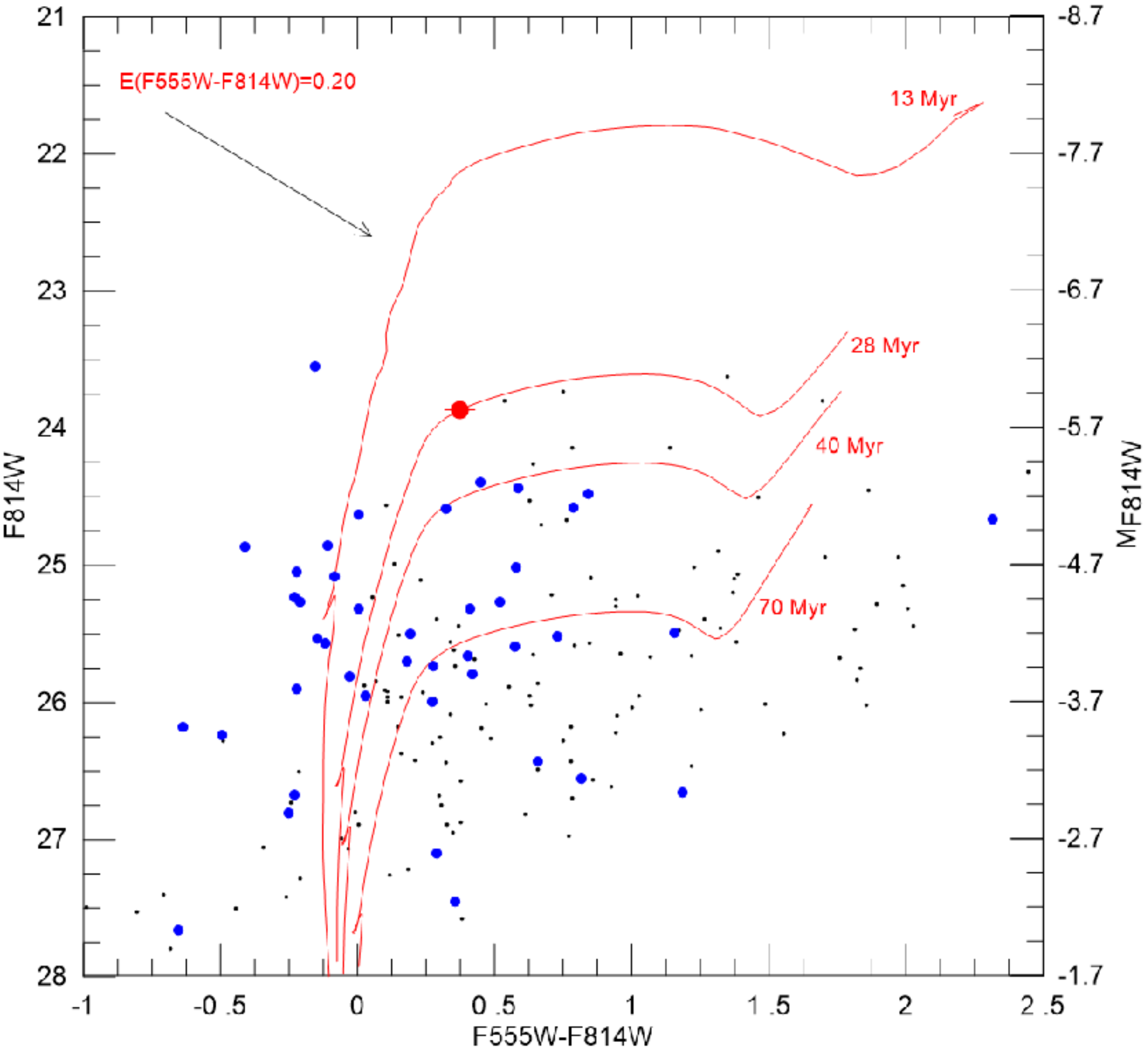}
\caption{The CMD for the possible counterpart, group of star and field stars around X-4. The red circle and black dots represent the possible optical counterpart of X-4 and the field stars within the 2\arcsec region around the source, respectively. The blue dots represent stars in the group. The isochrones have been corrected for extinction of $A_{V} =0.5$ mag and the black arrow shows the reddening line.}
\end{center}
\end{figure*}

\begin{figure*}
\label{Fig2}
\begin{center}
\includegraphics[scale=0.27]{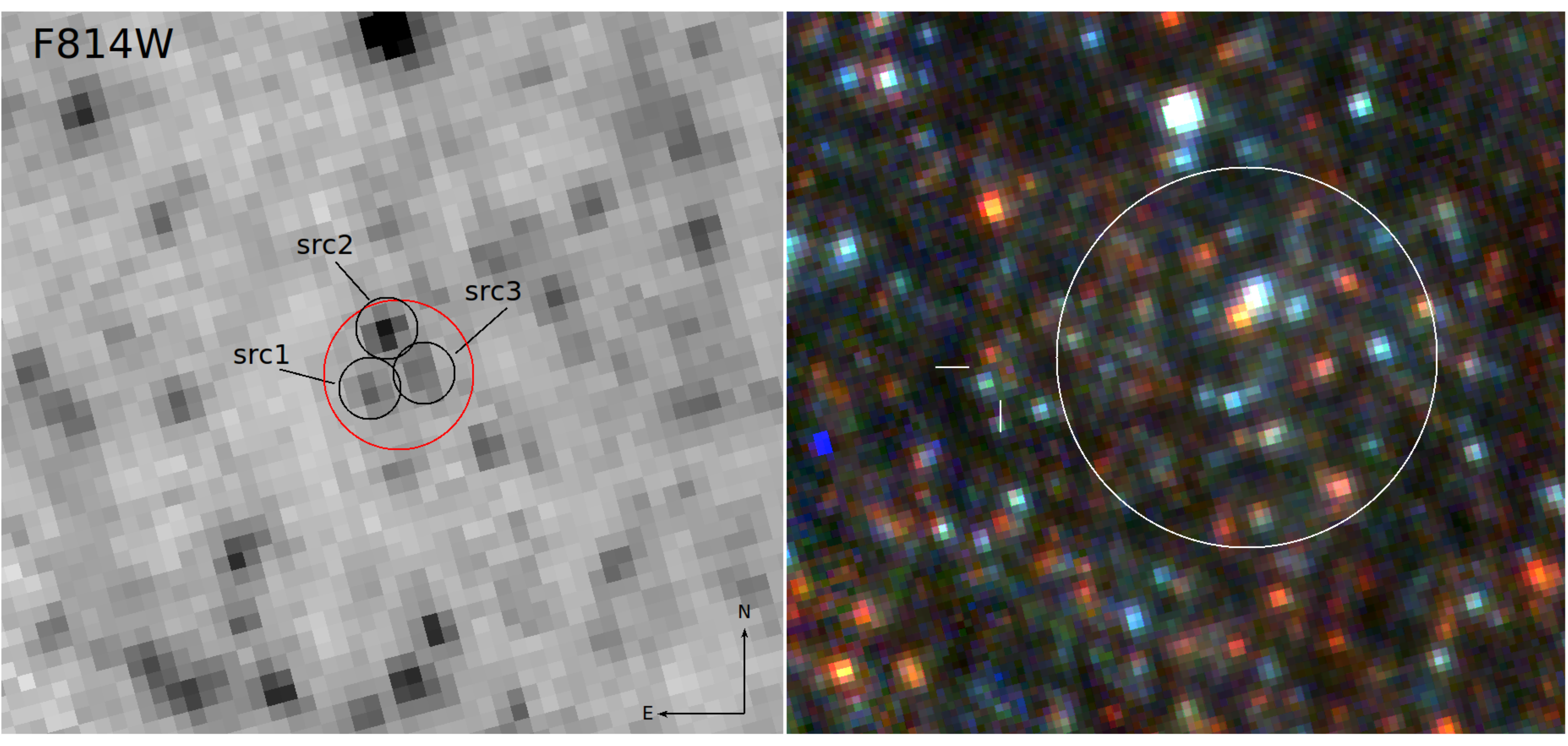}
\caption{The {\it HST}/WFC3 F814W image of X-6 (left) and the true color {\it HST} image of the star group near the source (Red: F814W, Green: F555W, Blue: F438W) (right). In the left panel, the red circle represents the corrected position of X-6 with an accuracy of 0\farcs19 radius. Three optical candidates (black circles, the aperture of PSF fitting is 0\farcs12 radius) are found within the error circle. The zoomed image has a size of $\sim 2\arcsec \times 1\farcs7$. In the right panel, white circle (2\arcsec diameter) represents the group of stars $\sim 1\farcs3$ away from X-6. The optical candidates are marked with white bars. The RGB image has a size of $\sim 4\farcs8 \times 4\farcs2$.}
\end{center}
\end{figure*}

\begin{figure*}
\label{Fig2}
\begin{center}
\includegraphics[scale=0.41]{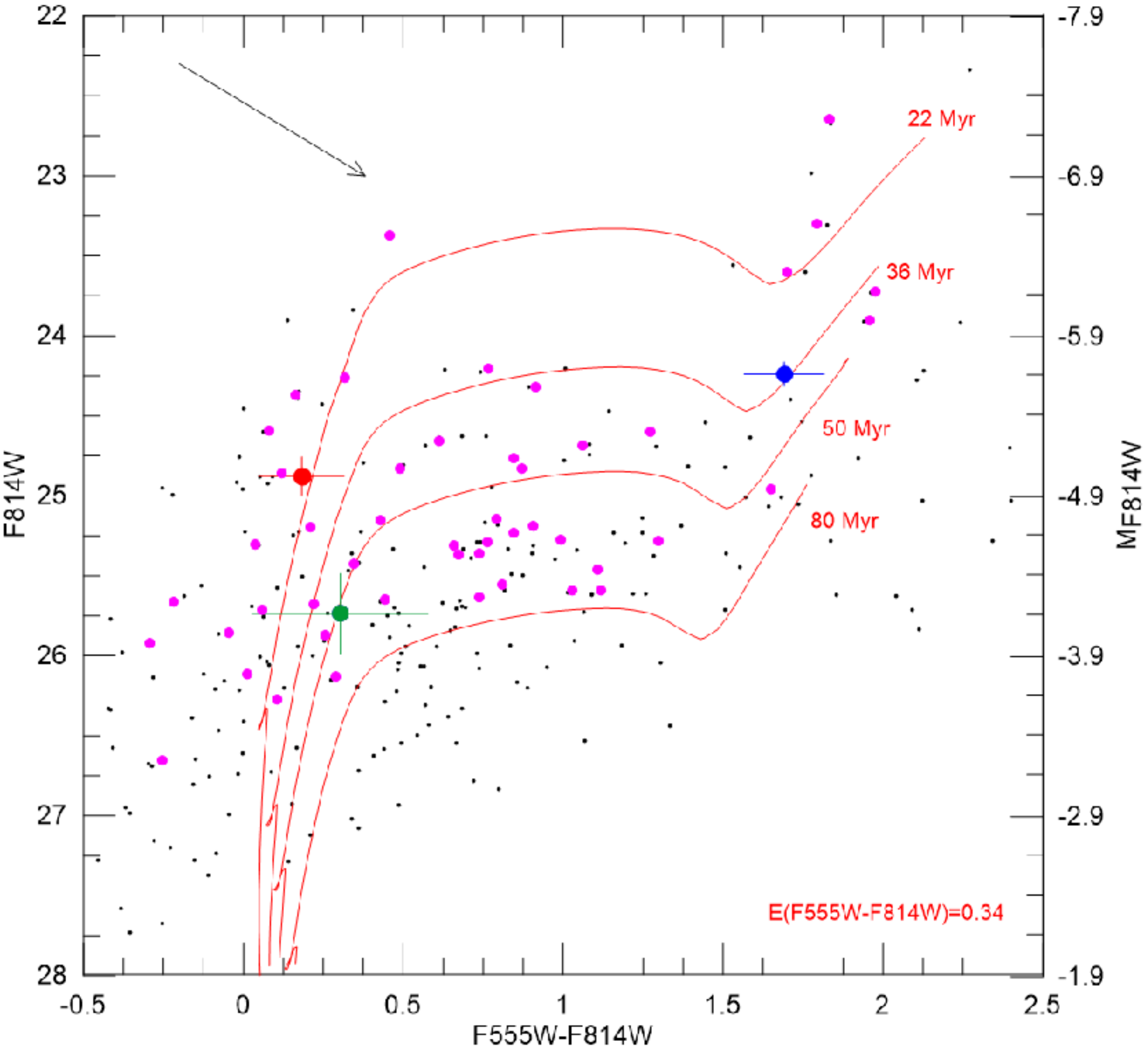}
\caption{The CMD for the possible counterparts, group of stars and field stars around X-6. The red, blue and green circles represent src1, src2 and src3 respectively. The black and magenta dots represent the field stars within the 2\arcsec region around X-6 and star groups, respectively. The isochrones have been corrected for extinction of $A_{V}=0.85$ mag and the black arrow shows the reddening line.}
\end{center}
\end{figure*}

\begin{figure*}
\label{Fig2}
\begin{center}
\includegraphics[scale=0.27]{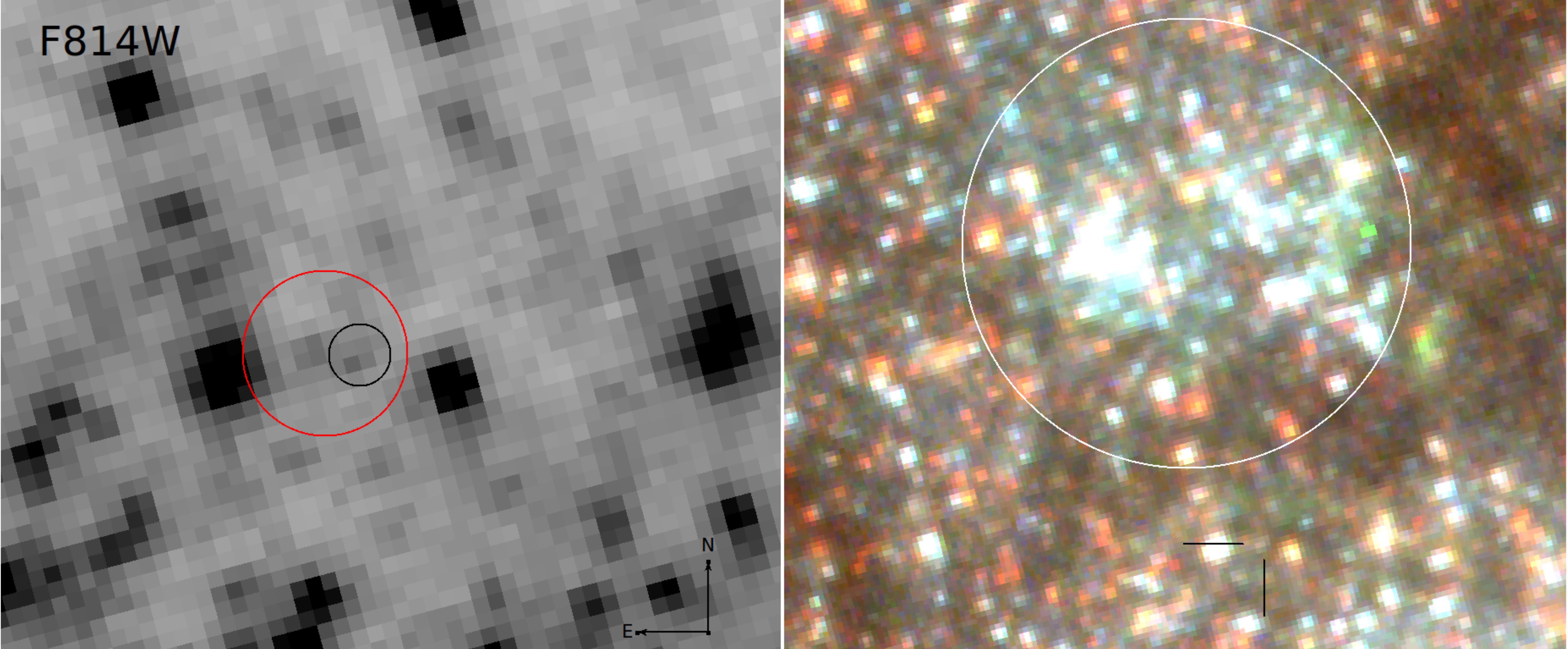}
\caption{The {\it HST}/WFC3 F814W image of X-7 (left) and the true color {\it HST} image of the star group near X-7 (Red: F814W, Green: F555W, Blue: F438W) (right). In the left panel, the red circle represents the corrected position of X-7 with an accuracy of 0\farcs21 radius. Although two sources are seen within this radius, the fainter source(m$=$ 26.7$\pm0.13$) located in the center of the red circle was not identified in other filters. Remaning possible optical counterpart (black circle, the aperture of PSF fitting is 0\farcs12 radius) is determined as a candidate. The zoomed image has a size of $\sim 1\farcs6 \times 1\farcs3$. In the right panel, white circle (3\farcs4  diameter) represents the group of stars $\sim 2\farcs5$ away from  X-7. The optical candidate is marked with black bars. The RGB image has a size of $\sim 8\farcs5 \times 4\arcsec$.}
\end{center}
\end{figure*}

 \begin{figure*}
\label{Fig2}
\begin{center}
\includegraphics[scale=0.41]{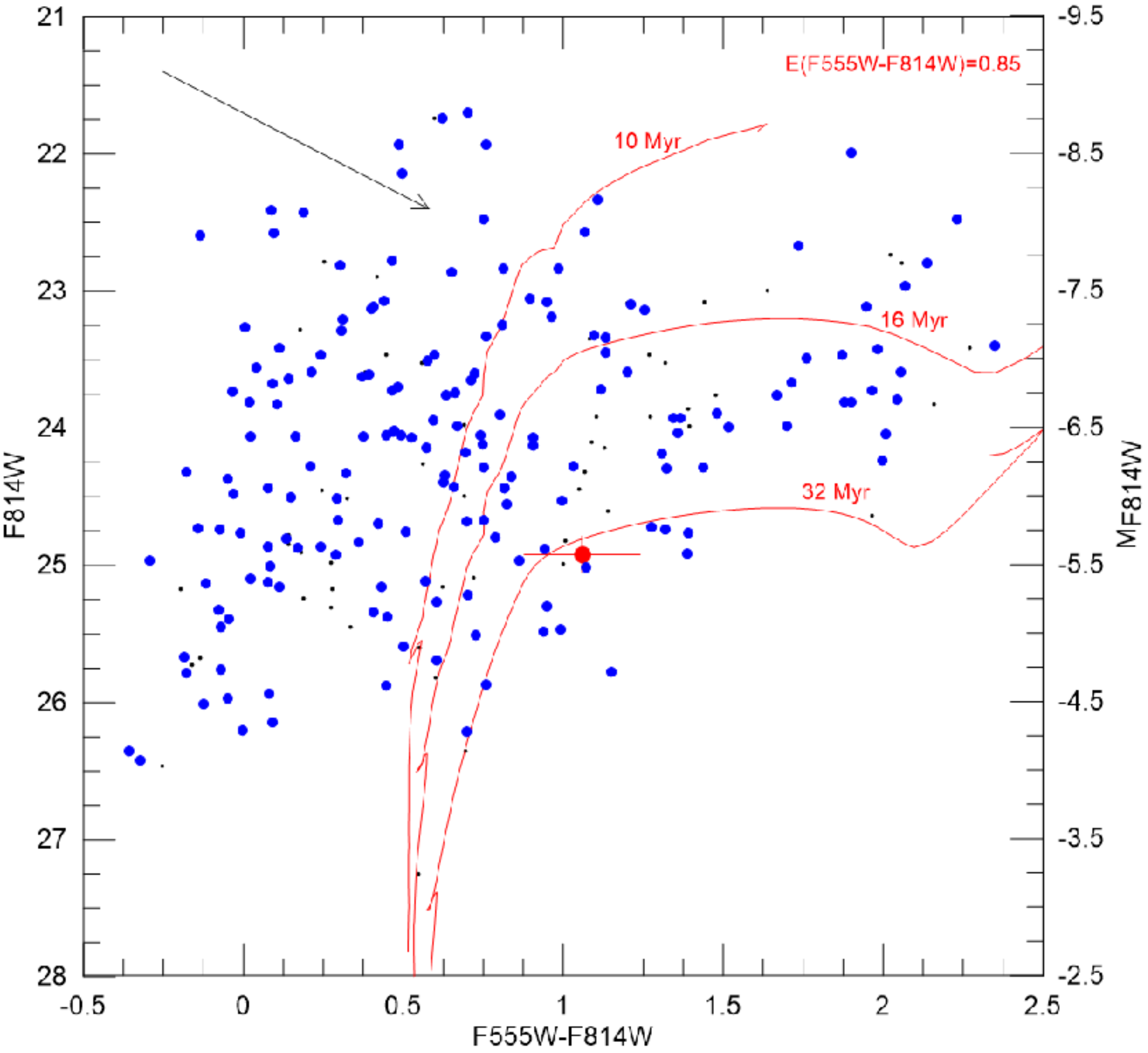}
\caption{The CMD for the possible counterpart, group of stars and field stars around X-7. The red circle represents possible counterpart of X-7. The black and blue dots represent the field stars within 3\arcsec region around the source and stars in gruop, respectively. The isochrones have been corrected for extinction of $A_{V}=2.1$ mag and the black arrow shows the reddening line.}
\end{center}
\end{figure*}

 \begin{figure*}
\label{Fig2}
\begin{center}
\includegraphics[scale=0.7]{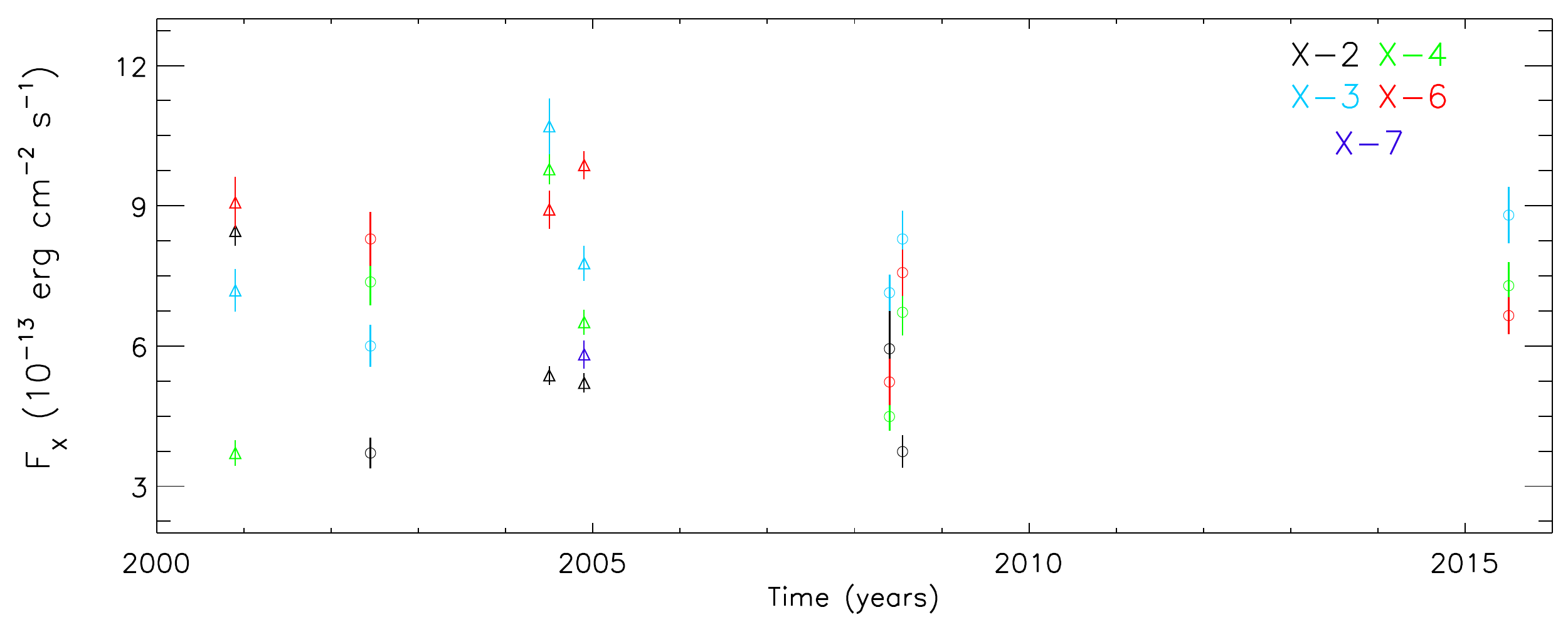}
\caption{Long-term light curves for X-2, X-3, X-4, X-6 and X-7. Triangles and circles represent {\it Chandra} and {\it XMM-Newton} data, respectively. The flux values are unabsorbed and calculated in the 0.5$-$8 keV energy band.}
\end{center}
\end{figure*}

 \begin{figure*}
\label{Fig2}
\begin{center}
\includegraphics[scale=0.2]{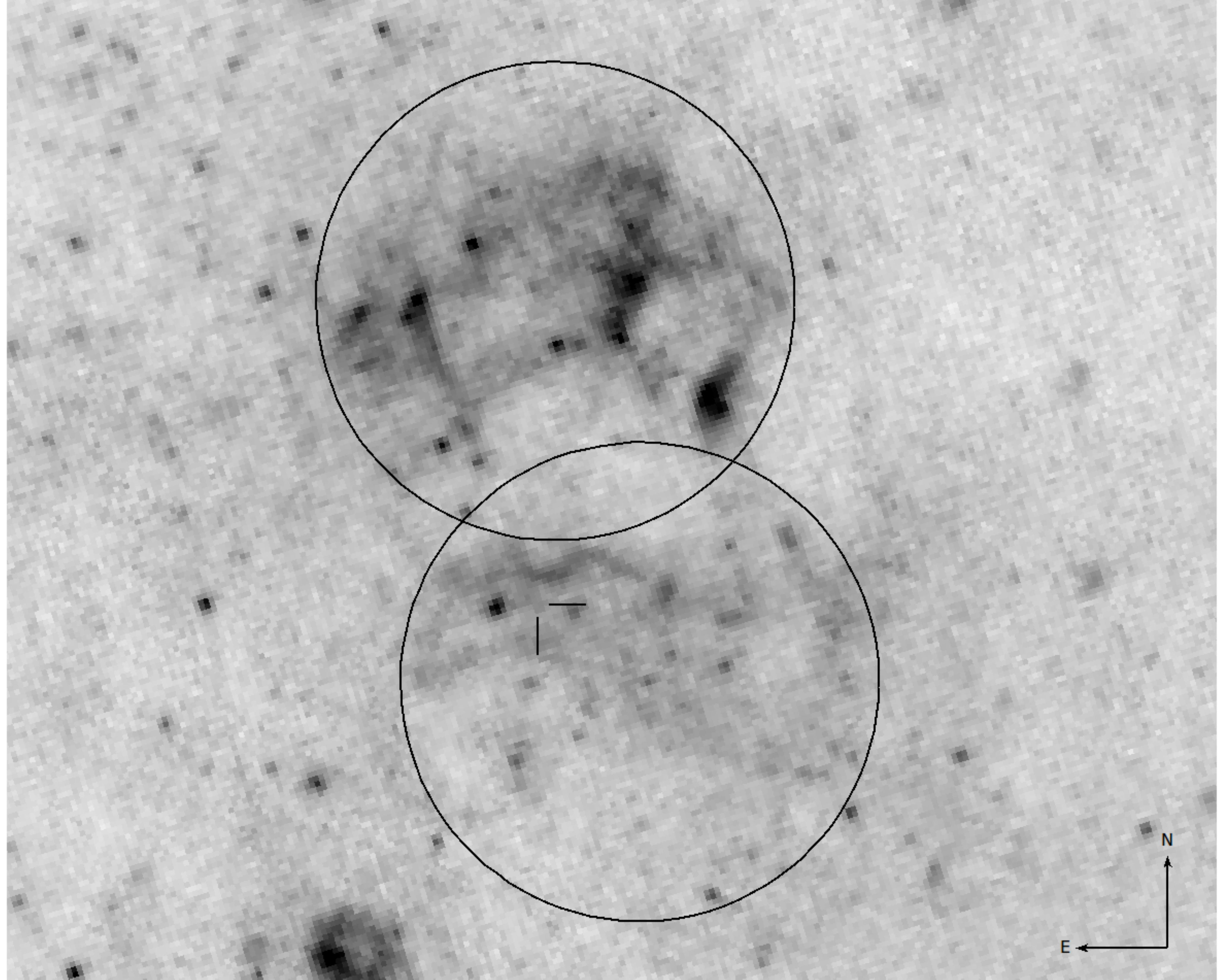}
\caption{{\it  HST}/WFC3 F657N filter image shows $\mathrm{H}\alpha$  nebula around the ULX X-7. The black circles (radius of 1\farcs8) represent north and south parts of nebula and black bars indicate the position of X-7.}
\end{center}
\end{figure*}


\begin{thebibliography}{}

\bibitem[Abolmasov et al.(2007)]{2007ApJ...668..124A} Abolmasov, P.~K., Swartz, D.~A., Fabrika, S., et al.\ 2007a, \apj, 668, 124
\bibitem[Abolmasov et al.(2007)]{2007AstBu..62...36A} Abolmasov, P., Fabrika, S., Sholukhova, O., \& Afanasiev, V.\ 2007b, Astrophysical Bulletin, 62, 36
\bibitem[Abolmasov et al.(2008)]{2008arXiv0809.0409A} Abolmasov, P., Fabrika, S., Sholukhova, O., \& Kotani, T.\ 2008, arXiv:0809.0409 
\bibitem[Afanasiev \& Moiseev(2005)]{2005AstL...31..194A} Afanasiev, V.~L., \& Moiseev, A.~V.\ 2005, Astronomy Letters, 31, 194
\bibitem[Aird et al.(2010)]{2010MNRAS.401.2531A} Aird, J., Nandra, K., Laird, E.~S., et al.\ 2010, \mnras, 401, 2531
\bibitem[Alam et al.(2015)]{2015ApJS..219...12A} Alam, S., Albareti, F.~D., Allende Prieto, C., et al.\ 2015, \apjs, 219, 12
\bibitem[Aller et al.(1982)]{1982lbg6.conf.....A} Aller, L.~H., Appenzeller, I., Baschek, B., et al.\ 1982, Landolt-Bornstein: Group 6: Astronomy, 54
\bibitem[Avdan et al.(2016)]{2016ApJ...828..105A} Avdan, H., Avdan, S., Akyuz, A., et al.\ 2016a, \apj, 828, 105
\bibitem[Avdan et al.(2016)]{2016MNRAS.455L..91A} Avdan, S., Vinokurov, A., Fabrika, S., et al.\ 2016b, \mnras, 455, L91
\bibitem[Bachetti et al.(2014)]{2014Natur.514..202B} Bachetti, M., Harrison, F.~A., Walton, D.~J., et al.\ 2014, \nat, 514, 202
\bibitem[Bressan et al.(2012)]{2012MNRAS.427..127B} Bressan, A., Marigo, P., Girardi, L., et al.\ 2012, \mnras, 427, 127
\bibitem[Calzetti et al.(1994)]{1994ApJ...429..582C} Calzetti, D., Kinney, A.~L., \& Storchi-Bergmann, T.\ 1994, \apj, 429, 582
\bibitem[Calzetti(2001)]{2001PASP..113.1449C} Calzetti, D.\ 2001, \pasp, 113, 1449  
\bibitem[Cardelli et al.(1989)]{1989ApJ...345..245C} Cardelli, J.~A., Clayton, G.~C., \& Mathis, J.~S.\ 1989, \apj, 345, 245
\bibitem[Carpano et al.(2018)]{2018MNRAS.476L..45C} Carpano, S., Haberl, F., Maitra, C., \& Vasilopoulos, G.\ 2018, \mnras, 476, L45 
\bibitem[Clemens \& Alexander(2002)]{2002MNRAS.333...39C} Clemens, M.~S., \& Alexander, P.\ 2002, \mnras, 333, 39
\bibitem[Clemens et al.(1999)]{1999MNRAS.307..481C} Clemens, M.~S., Alexander, P., \& Green, D.~A.\ 1999, \mnras, 307, 481
\bibitem[Copperwheat(2007)]{2007PhDT.......298C} Copperwheat, C.~M.\ 2007, Ph.D.~Thesis
\bibitem[Della Ceca et al.(2015)]{2015MNRAS.447.3227D} Della Ceca, R., Carrera, F.~J., Caccianiga, A., et al.\ 2015, \mnras, 447, 3227
\bibitem[Dolphin(2000)]{2000PASP..112.1383D} Dolphin, A.~E.\ 2000, \pasp, 112, 1383
\bibitem[Dowell et al.(2008)]{2008AJ....135..823D} Dowell, J.~D., Buckalew, B.~A., \& Tan, J.~C.\ 2008, \aj, 135, 823
\bibitem[Esposito et al.(2013)]{2013MNRAS.436.3380E} Esposito, P., Israel, G.~L., Sidoli, L., et al.\ 2013, \mnras, 436, 3380
\bibitem[F{\"u}rst et al.(2016)]{2016ApJ...831L..14F} F{\"u}rst, F., Walton, D.~J., Harrison, F.~A., et al.\ 2016, \apjl, 831, L14
\bibitem[Fabrika et al.(2015)]{2015NatPh..11..551F} Fabrika, S., Ueda, Y., Vinokurov, A., Sholukhova, O., \& Shidatsu, M.\ 2015, Nature Physics, 11, 551
\bibitem[Feng \& Kaaret(2008)]{2008ApJ...675.1067F} Feng, H., \& Kaaret, P.\ 2008, \apj, 675, 1067 
\bibitem[Feng \& Soria(2011)]{2011NewAR..55..166F} Feng, H., \& Soria, R.\ 2011, \nar, 55, 166
\bibitem[Fridriksson et al.(2008)]{2008ApJS..177..465F} Fridriksson, J.~K., Homan, J., Lewin, W.~H.~G., Kong, A.~K.~H., \& Pooley, D.\ 2008, \apjs, 177, 465
\bibitem[Gladstone \& Roberts(2009)]{2009MNRAS.397..124G} Gladstone, J.~C., \& Roberts, T.~P.\ 2009, \mnras, 397, 124
\bibitem[Gris{\'e} et al.(2012)]{2012ApJ...745..123G} Gris{\'e}, F., Kaaret, P., Corbel, S., et al.\ 2012, \apj, 745, 123
\bibitem[Gris{\'e} et al.(2011)]{2011ApJ...734...23G} Gris{\'e}, F., Kaaret, P., Pakull, M.~W., \& Motch, C.\ 2011, \apj, 734, 23
\bibitem[Gris{\'e} et al.(2008)]{2008A&A...486..151G} Gris{\'e}, F., Pakull, M.~W., Soria, R., et al.\ 2008, \aap, 486, 151
\bibitem[Gris{\'e} et al.(2005)]{2005sf2a.conf..549G} Gris{\'e}, F., Pakull, M., \& Motch, C.\ 2005, SF2A-2005: Semaine de l'Astrophysique Francaise, 549
\bibitem[Heida et al.(2016)]{2016MNRAS.459..771H} Heida, M., Jonker, P.~G., Torres, M.~A.~P., et al.\ 2016, \mnras, 459, 771
\bibitem[Heida et al.(2014)]{2014MNRAS.442.1054H} Heida, M., Jonker, P.~G., Torres, M.~A.~P., et al.\ 2014, \mnras, 442, 1054 
\bibitem[Israel et al.(2017)]{2017Sci...355..817I} Israel, G.~L., Belfiore, A., Stella, L., et al.\ 2017a, Science, 355, 817
\bibitem[Israel et al.(2017)]{2017MNRAS.466L..48I} Israel, G.~L., Papitto, A., Esposito, P., et al.\ 2017b, \mnras, 466, L48
\bibitem[Jonker et al.(2012)]{2012ApJ...758...28J} Jonker, P.~G., Heida, M., Torres, M.~A.~P., et al.\ 2012, \apj, 758, 28
\bibitem[Kaaret et al.(2004)]{2004MNRAS.351L..83K} Kaaret, P., Ward, M.~J., \& Zezas, A.\ 2004, \mnras, 351, L83
\bibitem[Kaaret(2005)]{2005ApJ...629..233K} Kaaret, P.\ 2005, \apj, 629, 233 
\bibitem[Kaaret et al.(2017)]{2017ARA&A..55..303K} Kaaret, P., Feng, H., \& Roberts, T.~P.\ 2017, \araa, 55, 303
\bibitem[Kubota et al.(1998)]{kub98} Kubota, A., Tanaka, Y., Makishima, K., et al. 1998, \pasj, 50, 667
\bibitem[Lehmann et al.(2005)]{2005A&A...431..847L} Lehmann, I., Becker, T., Fabrika, S., et al.\ 2005, \aap, 431, 847 
\bibitem[Liu et al.(2004)]{2004ApJ...602..249L} Liu, J.-F., Bregman, J.~N., \& Seitzer, P.\ 2004, \apj, 602, 249
\bibitem[Liu et al.(2007)]{2007ApJ...661..165L} Liu, J.-F., Bregman, J., Miller, J., \& Kaaret, P.\ 2007, \apj, 661, 165
\bibitem[Liu et al.(2013)]{2013Natur.503..500L} Liu, J.-F., Bregman, J.~N., Bai, Y., Justham, S., \& Crowther, P.\ 2013, \nat, 503, 500
\bibitem[L{\'o}pez et al.(2017)]{2017MNRAS.469..671L} L{\'o}pez, K.~M., Heida, M., Jonker, P.~G., et al.\ 2017, \mnras, 469, 671 
\bibitem[Makishima et al.(2000)]{mak00} Makishima, K., Kubota, A., Mizuno, T., et al. 2000, \apj, 535, 632
\bibitem[Motch et al.(2014)]{2014Natur.514..198M} Motch, C., Pakull, M.~W., Soria, R., Gris{\'e}, F., \& Pietrzy{\'n}ski, G.\ 2014, \nat, 514, 198
\bibitem[Osterbrock(1989)]{1989agna.book.....O} Osterbrock, D.~E.\ 1989, Research supported by the University of California, John Simon Guggenheim Memorial Foundation, University of Minnesota, et al.~Mill Valley, CA, University Science Books, 1989, 422 p.
\bibitem[Pandey et al.(2003)]{2003A&A...397..191P} Pandey, A.~K., Upadhyay, K., Nakada, Y., \& Ogura, K.\ 2003, \aap, 397, 191
\bibitem[Patruno \& Zampieri(2010)]{2010MNRAS.403L..69P} Patruno, A., \& Zampieri, L.\ 2010, \mnras, 403, L69 
\bibitem[Patruno \& Zampieri(2008)]{2008MNRAS.386..543P} Patruno, A., \& Zampieri, L.\ 2008, \mnras, 386, 543
\bibitem[Poutanen et al.(2013)]{2013MNRAS.432..506P} Poutanen, J., Fabrika, S., Valeev, A.~F., Sholukhova, O., \& Greiner, J.\ 2013, \mnras, 432, 506
\bibitem[Roberts et al.(2008)]{2008MNRAS.387...73R} Roberts, T.~P., Levan, A.~J., \& Goad, M.~R.\ 2008, \mnras, 387, 73
\bibitem[Roberts \& Warwick(2000)]{2000MNRAS.315...98R} Roberts, T.~P., \& Warwick, R.~S.\ 2000, \mnras, 315, 98
\bibitem[Roberts et al.(2002)]{2002MNRAS.337..677R} Roberts, T.~P., Warwick, R.~S., Ward, M.~J., \& Murray, S.~S.\ 2002, \mnras, 337, 677
\bibitem[Shimura \& Takahara(1995)]{shi95} Shimura, T., \& Takahara, F. \apj, 445, 780
\bibitem[Soria et al.(2005)]{2005MNRAS.356...12S} Soria, R., Cropper, M., Pakull, M., Mushotzky, R., \& Wu, K.\ 2005, \mnras, 356, 12
\bibitem[Straizys (1995)]{stra95} Straizys, V. 1995, Multicolor Stellar Photometry, Astronomy and Astrophysics Series, Vol. 15, ed. A. G. Pacholczyk (Tucson, Arizona: Pachart Pub. House)
\bibitem[Strauss et al.(1992)]{1992ApJS...83...29S} Strauss, M.~A., Huchra, J.~P., Davis, M., et al.\ 1992, \apjs, 83, 29
\bibitem[Swartz et al.(2011)]{2011ApJ...741...49S} Swartz, D.~A., Soria, R., Tennant, A.~F., \& Yukita, M.\ 2011, \apj, 741, 49
\bibitem[Tao et al.(2011)]{2011ApJ...737...81T} Tao, L., Feng, H., Gris{\'e}, F., \& Kaaret, P.\ 2011, \apj, 737, 81 
\bibitem[Tully et al.(1998)]{1998AJ....115.2264T} Tully, R.~B., Pierce, M.~J., Huang, J.-S., et al.\ 1998, \aj, 115, 2264
\bibitem[van Paradijs(1981)]{1981A&A...103..140V} van Paradijs, J.\ 1981, \aap, 103, 140
\bibitem[Vinokurov et al.(2018)]{2018ApJ...854..176V} Vinokurov, A., Fabrika, S., \& Atapin, K.\ 2018, \apj, 854, 176
\bibitem[Yang et al.(2011)]{2011ApJ...733..118Y} Yang, L., Feng, H., \& Kaaret, P.\ 2011, \apj, 733, 118
\bibitem[Yoshida et al.(2010)]{2010ApJ...722..760Y} Yoshida, T., Ebisawa, K., Matsushita, K., Tsujimoto, M., \& Kawaguchi, T.\ 2010, \apj, 722, 760

\end{thebibliography}
\end{document}